\documentclass[12pt]{article}
\usepackage{graphicx}
\usepackage{amsmath,amssymb,amsthm}
\usepackage{booktabs}
\usepackage{array}
\usepackage{pdfpages}
\usepackage{setspace}
\usepackage{natbib}
\usepackage{color}
\usepackage{geometry}
\newtheorem{assumption}{Assumption}
\newtheorem{theorem}{Theorem}

\newtheorem{proposition}{Proposition}
\allowdisplaybreaks[4]

\title{\bf Difference-in-Differences with a Mediator}
\author{Yuhao Deng$^1$, Haoyu Wei$^2$ and Zhongzhe Ouyang$^{3*}$ \\
{\small 1 Fred Hutchinson Cancer Center} \\
{\small 2 University of California (San Diego)} \\
{\small 3 Guangzhou University}}
\date{}

\begin{document}

\maketitle

\begin{abstract}
Causal mediation analysis is a powerful tool for decomposing the total effect of a treatment into its direct effect on the outcome and its indirect effect mediated through an intermediate variable. However, in observational studies, confounding between treatment and potential outcomes typically renders the total and natural effects non-identifiable. In this work, we advance the mediation analysis within the difference-in-differences framework. Under a mediator-adjusted parallel trends assumption and additional conditions, we demonstrate that natural indirect, direct, and total effects are identifiable in the treated group. We further derive efficient influence functions for these estimands, enabling the construction of multiply robust and nonparametrically efficient estimators. Applying our methodology to data from the Job Corps Study, we find that job training has a positive direct effect on short-term earnings, holding employment intensity fixed at its level under training, which is masked in the total effect by reduced work time; in the longer run, the total effect is positive and operates mainly through higher employment intensity. \par
Keywords: Mediation analysis; natural effect; controlled effect; efficient influence function; multiple robustness; Job Corps
\end{abstract}

\section{Introduction}

In the economic and health sciences, studying the mechanisms by which treatment affects outcomes has been both interesting and challenging. Treatment can affect outcomes either directly or indirectly through a mediator. Mediation analysis was initially proposed within linear models \citep{baron1986moderator}. Later, mediation analysis was formalized using potential outcomes, in which the mediator is a potential outcome of the treatment and the primary outcome is a potential outcome of the treatment and the mediator \citep{rubin2004direct, goetgeluk2008estimation}. By controlling for the mediator at the natural level, the total effect decomposes into natural indirect and direct effects. Identifiability assumptions have been proposed to identify the natural indirect and direct effects \citep{joffe2009related, imai2010general, imai2010identification}. In general, mediation analysis methods require that the treatment assignment, mediator, and primary outcome be unconfounded, which is a stringent condition in observational studies.

As a quasi-experimental design, difference-in-differences (DiD) was proposed to estimate treatment effects when the treatment is confounded with the outcome using panel data \citep{heckman1997matching, abadie2005semiparametric}. By leveraging pre-treatment outcomes, the change in potential outcomes from pre- to post-treatment is assumed to follow a common time trend across treatment groups. The pre-treatment outcome serves as a negative control variable for the unmeasured confounding \citep{zivich2023introducing}. Although the average treatment effect on the entire population is not identifiable, the mean potential outcome under control can be imputed for the treated units, so the average treatment effect in the treated group is. Doubly robust and locally efficient estimators of this effect have been derived from its efficient influence function \citep{chang2020double, sant2020doubly, callaway2021difference, deng2025improved}. 

Conducting mediation analysis in a difference-in-differences design can help address non-randomized treatment assignment. Existing work has followed several distinct identification strategies. \citet{deuchert2019direct} decompose the total effect into natural indirect and direct effects in experiments with non-compliance, assuming a monotonic effect of treatment on the binary mediator and imposing a parallel trends assumption within principal strata. \citet{huber2022direct} propose an alternative strategy based on changes-in-changes, which replaces parallel trends by distributional restrictions for binary mediators and continuous outcomes. Closest to our work, \citet{hsia2025causal} replaced no-unmeasured-confounding with parallel trends assumptions to identify population-level natural effects; identification is stated for a binary mediator conditional on the observed mediator, and a continuous mediator is accommodated only through a linear, additive two-way fixed effects regression. \citet{blackwell2025estimating} considered efficient estimation of controlled direct effects with a discrete mediator, conditioning on the observed mediator under a no-modified-direct-effect assumption. \citet{huber2026difference} identified treatment effects conditional on the observed post-treatment mediator in the treated group and defined natural effects by integration. Conditioning on the realized mediator makes trend assumptions examinable within mediator strata but potentially exposes them to collider bias. Section \ref{sec:compare} compares these strategies with ours.

Motivated by the Job Corps Study, we aim to study the effect of the job training program on earnings \citep{schochet2001national}. Although the assignment was randomized, the actual treatment was not, because participants could decide whether to join training programs regardless of their treatment assignment. Some previous work used instrumental variables or principal stratification to account for non-compliance; other work constructed bounds for the treatment effect \citep{schochet2008does, lee2009training, zhang2009likelihood, flores2010learning, flores2012estimating, blanco2013bounds, chen2015bounds}. These analyses indicated that job training has a significant long-term effect but an insignificant short-term effect on earnings. This may be because training takes up a large share of time, so wages are lower than those of full-time employees. Therefore, it is essential to quantify the treatment effect mediated by work time while accounting for non-randomized treatment.

In this paper, we consider the identification and estimation of mediation effects in two-period difference-in-differences. Our contributions are twofold. At the identification level, we show that natural direct and indirect effects in the treated group are identified under a parallel trends assumption, which conditions on the counterfactual mediator, and a sequential ignorability condition on trends. These assumptions allow for unmeasured confounding between treatment and outcome. At the estimation level, we derive efficient influence functions for the natural and total effects and construct multiply robust, locally efficient estimators. For controlled effects with a continuous mediator, where pathwise differentiability of the estimand fails \citep{kennedy2017non}, we use kernel smoothing and show how to handle mediators with a mixed distribution. In the Job Corps Study, we find that training has a positive direct effect on short-term earnings holding employment intensity at its level under training, and that its positive long-run effect operates mainly through increased employment.

The remainder of this paper is organized as follows. Section \ref{sec:framework} introduces the framework, defines the estimands, and establishes identifiability. Section \ref{sec:estimation} develops estimators based on efficient influence functions, and Section \ref{sec:control} extends them to controlled effects. Section \ref{sec:simulation} reports simulation studies, Section \ref{sec:application} analyzes the Job Corps data, and Section \ref{sec:discussion} concludes.

\section{Framework} \label{sec:framework}

\subsection{Estimands: natural indirect and direct effects}

In a two-group and two-period setting, let $G \in \{0,1\}$ be the treatment group assignment and $t \in \{0,1\}$ be the period indicator. All units are unexposed in the pre-treatment period, and only units in the treated group are exposed in the post-treatment period, so the treatment received in period $t$ is $D_t=Gt$. Throughout, we use $G$ to denote both the group and the post-treatment exposure.
Let $Y_0(g)$ denote the potential outcome in the pre-treatment period under treatment assignment $g$. Suppose there is a mediator that may mediate the treatment effect on the outcome. Let $M(g)$ denote the potential mediator \emph{in the post-treatment period} under treatment assignment $g$, shorthand for $M_1(g)$. The mediator can be either discrete or continuous. Let $Y_1(g,m)$ denote the potential outcome in the post-treatment period under treatment assignment $g$ given the mediator $m$. When the mediator is at the natural value $M(g)$ under treatment assignment $g$, the potential outcome in the post-treatment period is $Y_1(g,M(g))$, which we denote as $Y_1(g)$. In some cases, the mediator can also be measured in the pre-treatment period; we denote the pre-treatment potential mediator by $M_0(g)$ and the pre-treatment potential outcome by $Y_0(g,m)$.

A key question in mediation analysis is the extent to which the treatment effect is mediated by the mediator. In our setting, treatment may not be randomized, so the distribution of unmeasured features may differ across treatment groups. Consistent with the difference-in-differences literature, we adopt the treated group $G=1$ as the target population to minimize the identification requirements. The total effect is 
\begin{align*}
\tau &= E\{Y_1(1) - Y_1(0) \mid G=1\} \\
&= E\{Y_1(1,M(1)) - Y_1(0,M(0)) \mid G=1\}.
\end{align*}
By switching the treatment associated with the mediator, we define the natural indirect effect (NIE) as 
\[
\tau_{IE} = E\{Y_1(0,M(1))-Y_1(0,M(0)) \mid G=1\}.
\]
By switching the treatment associated with the outcome, we define the natural direct effect (NDE) as 
\[
\tau_{DE} = E\{Y_1(1,M(1))-Y_1(0,M(1)) \mid G=1\}.
\]
The natural indirect effect measures the effect delivered through the mediator, and the natural direct effect measures the effect delivered without being mediated by the mediator, holding the mediator at the level it would take under treatment, $M(1)$. Accordingly, the total effect can be decomposed into the natural indirect and direct effects.

%The definitions of natural (in)direct effects are not unique. In our definition of the natural indirect effect, we fix the baseline treatment at $g=0$, and vary the mediator from $M(0)$ to $M(1)$. Correspondingly, the natural direct effect is defined by fixing the mediator at $M(1)$ and switching the baseline treatment from 0 to 1. We can also define the natural indirect effect by controlling the baseline treatment at $g=1$ and then switching the mediator. While such a definition is mathematically valid, it may lack interpretability in practice. In our motivating data, the treatment corresponds to training, and the mediator is the proportion of weeks employed. Since training occupies time, $M(1)$ is generally smaller than $M(0)$. As a result, the potential outcome $Y_1(1,M(0))$ may be infeasible, as an individual cannot participate in training and work simultaneously.

\subsection{Identifiability}

Let $X$ be baseline covariates measured prior to treatment, which may include the pre-treatment mediator $M_0$. Let $Y_0$ and $Y_1$ be the observed outcomes in the pre-treatment and post-treatment periods. The observed data consist of $n$ independent and identically distributed copies of $O=(G,X,Y_0,M,Y_1)$. To identify the natural indirect, direct, and total effects for the treated group, we adopt a difference-in-differences framework and impose the following assumptions.

\begin{assumption}[No anticipation] \label{ass1}
(i) $Y_0(g) = Y_0(0)$ for $g=0,1$; (ii) $Y_0(g,m) = Y_0(0)$ for all $g$ and $m$; (iii) $M_0(1) = M_0(0)$.
\end{assumption}

Assumption \ref{ass1} rules out anticipation effects: neither future treatment (i) nor a future intervention on the mediator (ii) affects the pre-treatment outcome, and the pre-treatment mediator is unaffected by future treatment (iii). Parts (i) and (ii) say that potential outcomes depend on the treatment actually received. Part (iii) is needed whenever $M_0$ is included in $X$; if $M_0$ were affected by anticipated treatment, conditioning on it would introduce bias. These conditions parallel those of \citet{hsia2025causal}.

\begin{assumption}[Mediator-adjusted parallel trends] \label{ass2}
For all $m$, $E\{Y_1(0,m)-Y_0(0) \mid M(0)=m, G=1, X\} = E\{Y_1(0,m)-Y_0(0) \mid M(0)=m, G=0, X\}$.
\end{assumption}

Assumption \ref{ass2} extends the standard parallel trends assumption to accommodate a mediator: within latent strata defined by the value $m$ that the mediator would take in the absence of treatment, and conditional on baseline covariates, the time trend in untreated potential outcomes is identical across groups. Treatment may be confounded with the level of potential outcomes; differencing over time removes time-invariant unobserved confounding.

The conditioning event involves the counterfactual mediator $M(0)$, which is never observed in the treated group. The benefit is that we do not condition on a post-treatment variable, thereby avoiding the collider bias that may arise when outcome trends are compared across groups within levels of the realized mediator \citep{blackwell2025estimating, huber2026difference}. The cost is that Assumption \ref{ass2} is a cross-world restriction that a pre-trend analysis cannot examine directly, even with many pre-treatment periods, because treated units cannot be stratified by $M(0)$. We thus purchase freedom from collider bias at the price of the partial verifiability that makes the standard parallel trends assumption persuasive in applied work; the same limitation applies to \citet{hsia2025causal}. What remains testable is formalized in Proposition \ref{prop:pt} later.

\begin{assumption}[Sequential ignorability] \label{ass3}
(i) $G\perp M(g)\mid X$ for $g=0,1$; (ii) for all $m$,
$E\{Y_1(0,m)-Y_0(0) \mid M(1)=m, G=1, X\} = E\{Y_1(0,m)-Y_0(0) \mid M(0)=m, G=1, X\}$.
\end{assumption}

Part (i) requires no unmeasured confounding between treatment and the potential mediators given $X$. Part (ii) requires that, within the treated group, the untreated outcome trend at mediator level $m$ be the same for units whose mediator would equal $m$ under treatment and for units whose mediator would equal $m$ under control; that is, the mediator does not modify the underlying trends. This is weaker than classical sequential ignorability in two respects: it restricts conditional means rather than distributions, and it does not require $M(1)$ to be conditionally independent of $Y_1(1,m)$ in the treated group, so the mediator need not be as-if randomized. Two limitations should be stated with equal clarity. First, part (ii) is a cross-world condition, relating an outcome under $g=0$ to a mediator under $g=1$, and therefore cannot be verified from any observed data. Second, because $X$ consists of baseline covariates, Assumptions \ref{ass2} and \ref{ass3} exclude any treatment-induced confounder of the mediator--outcome relationship, even when it can be measured. When such confounding is suspected, randomized interventional effects serve as an alternative to natural effects \citep{vanderweele2017mediation}.

The distinct roles of Assumptions \ref{ass2} and \ref{ass3} are visible in the chain of equalities that underlies identifiability. In order to identify $E\{Y_1(0,M(1))-Y_0(0) \mid G=1, X\}$, Assumption \ref{ass3}(ii) moves from the $M(1)=m$ stratum to the $M(0)=m$ stratum within the treated group; Assumption \ref{ass2} transports the untreated trend in that stratum to the control group, where it is observed; and Assumption \ref{ass3}(i) lets the treated mediator distribution be read off the data.

\begin{proposition}[Testable implication] \label{prop:pt}
Under Assumption \ref{ass2} (for all $m$) and Assumption \ref{ass3}(i), the standard conditional parallel trends assumption holds: $E\{Y_1(0)-Y_0(0) \mid G=1, X\} = E\{Y_1(0)-Y_0(0) \mid G=0, X\}$.
\end{proposition}

Hence, standard parallel trends is not an additional requirement but an implication of our framework. With several pre-treatment periods, a conventional pre-trend test is therefore a legitimate partial falsification test: rejection implies that Assumption \ref{ass2} or \ref{ass3}(i) fails. However, passing the test does not validate the $M(0)$-conditional content of Assumption \ref{ass2} nor Assumption \ref{ass3}(ii).

\begin{assumption}[Positivity] \label{ass4}
$P(G=1)>0$ and $P(G=0 \mid M,X) > c$ almost surely for some constant $c>0$.
\end{assumption}

\begin{assumption}[Consistency] \label{ass5}
$M(G)=M$, $Y_0(G)=Y_0$ and $Y_1(G,M(G))=Y_1$.
\end{assumption}

Positivity ensures that the distributions of covariates and mediators in the treated group are dominated by those in the control group; $P(G=0\mid M,X)>c$ implies $P(G=0\mid X)>c$ by iterated expectations and, by Bayes' rule, $f(M\mid X)/f(M\mid G=0,X) = P(G=0\mid X)/P(G=0\mid M,X) \in (c, 1/c)$. Consistency links the potential outcomes and observed data. 

Let $\delta(g,m,x) = E(Y_1-Y_0 \mid G=g,M=m,X=x)$ denote the conditional mean of the change in observed outcomes. The following theorem expresses $\tau_{IE}$, $\tau_{DE}$, and $\tau$ in terms of this observable function.

\begin{theorem} \label{th:iden}
Under Assumptions \ref{ass1}--\ref{ass5}, the natural indirect, direct, and total effects are identifiable,
\begin{align*}
    \tau_{IE} = \tau(0,1) - \tau(0,0), \quad \tau_{DE} = \tau(1,1) - \tau(0,1), \quad \tau = \tau(1,1) - \tau(0,0),
\end{align*}
where $\tau(g,g^*) = E[E\{\delta(g,M,X) \mid G=g^*,X\} \mid G=1]$.
\end{theorem}
In $\tau(g,g^*)$, the argument $g$ is the level of treatment and $g^*$ the group whose mediator distribution is used. The inner expectation integrates $M$ out under the distribution of group $g^*$, and the outer expectation integrates $X$ out under the distribution of the treated group. Thus $\tau(1,1) = E\{\delta(1,M,X) \mid G=1\}$ is directly identifiable from the treated group, whereas $\tau(0,0)$ and $\tau(0,1)$ require transporting mediator and covariate distributions across groups. We define $\nu(g^*,X) = E\{\delta(0,M,X) \mid G=g^*,X\}$.

\subsection{When are the proposed assumptions plausible?}

\begin{figure}
    \centering
    \includegraphics[width=0.8\textwidth]{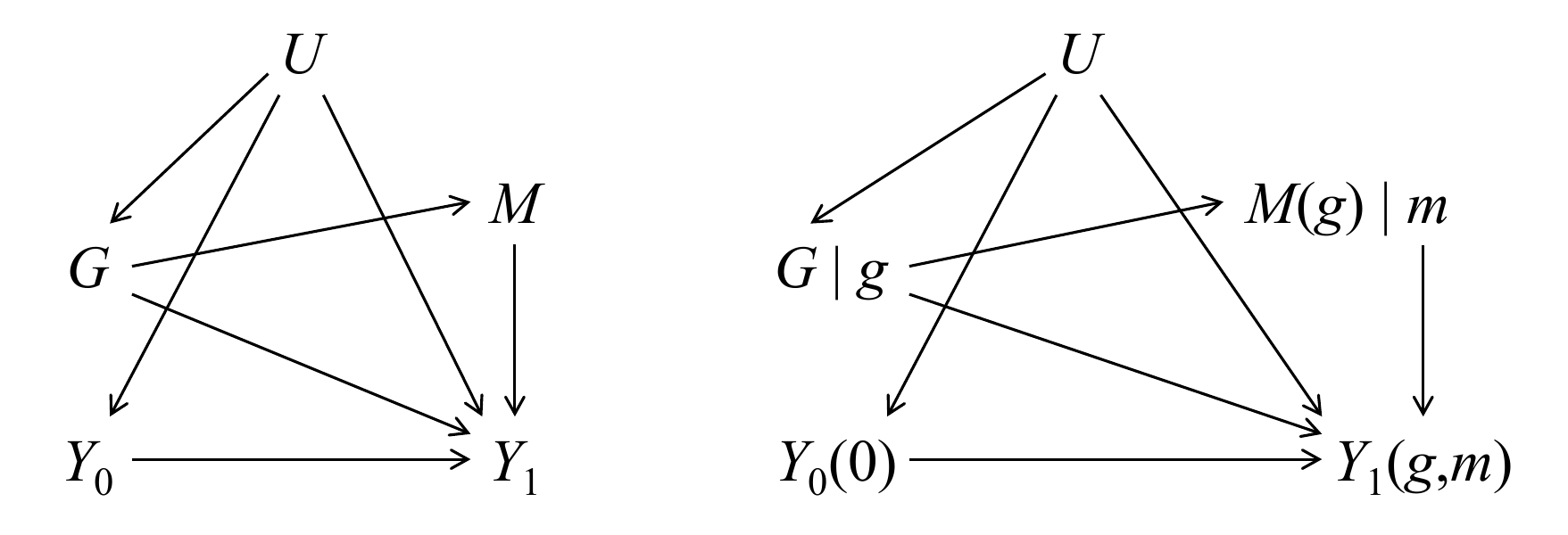}
    \caption{Directed acyclic graph (DAG) and single-world intervention graph (SWIG) of the mediator and primary outcomes in SCM 1. The intervention in $G$ introduces an exogenous variable $g$, and the intervention in $M(g)$ introduces an exogenous variable $m$ on the SWIG. Baseline covariates are omitted for simplicity.} \label{fig1}
\end{figure}

To clarify under which data-generating mechanisms the proposed assumptions are more plausible than conventional sequential ignorability, consider the structural causal model (SCM):
\begin{align*}
\mbox{SCM 1: } \quad G &= g(X,U) + \epsilon_G, \\
Y_0 &= f_0(X) + U + \epsilon_0, \\
M &= m(X,G) + \epsilon_M, \\
Y_1 &= f_1(X,G,M) + U + \epsilon_1,
\end{align*}
where $\epsilon_G$, $\epsilon_0$, $\epsilon_M$, and $\epsilon_1$ are mutually independent external errors, independent of $(X,U)$. The unmeasured confounder $U$ influences $G$, $Y_0$, and $Y_1$, but not $M$, so Assumption \ref{ass3}(i) holds. The change in potential outcomes under control, $Y_1(0,m)-Y_0(0) = f_1(X,0,m)-f_0(X)+\epsilon_1-\epsilon_0$, does not depend on $G$ (a function of $U$ and $\epsilon_G$), $M(0)$, or $M(1)$ (functions of $\epsilon_M$) given $X$, implying Assumptions \ref{ass2} and \ref{ass3}(ii). Conventional sequential ignorability fails because $U$ confounds $G$ and $Y_1$.
Figure \ref{fig1} shows the directed acyclic graph (DAG) and single-world intervention graph (SWIG): the mediator opens a ``front door'' between $G$ and $Y_1$ that is unaffected by $U$ \citep{pearl1995causal}. 

Another structural causal model that satisfies these assumptions is:
\begin{align*}
\mbox{SCM 2: } \quad G &= g(X) + \epsilon_G, \\
Y_0 &= f_0(X) + U + \epsilon_0, \\
M &= m(X,G,U) + \epsilon_M, \\
Y_1 &= f_1(X,G,M) + U + \epsilon_1.
\end{align*}
Here treatment is ignorable given $X$, but the mediator is influenced by the unmeasured $U$ that also shifts the level of both outcomes. Assumption \ref{ass3}(i) holds because $M(g)$ is independent of $\epsilon_G$ given $X$; Assumptions \ref{ass2} and \ref{ass3}(ii) hold because the trend does not involve $U$. Conventional sequential ignorability fails because $U$ confounds $M$ and $Y_1$.

In both SCMs, the unmeasured confounder enters the two outcomes additively and time-invariantly, so it cancels in the trend. What the assumptions still exclude is unmeasured confounding between treatment and mediator (in SCM 1, $U$ does not enter $m(\cdot)$) and between the mediator and the trend (in SCM 2, $U$ enters $m(\cdot)$ but not the trend). This assumption is plausible when selection into treatment is driven by persistent characteristics (ability, motivation, stable local labor-market conditions) and implausible when transitory shocks jointly affect take-up, the mediator, and the outcome change.

\subsection{Relation to existing identification strategies} \label{sec:compare}

Table \ref{tab:compare} compares the present framework with the mediation DiD literature. The closest framework is \citet{hsia2025causal}, which shares our core strategy, a cross-world condition, and the exclusion of treatment-induced confounding, but differs in three ways. First, it targets population-level effects and therefore requires parallel trends under both treatment and control, which the authors describe as implying homogeneous trends, whereas we target the treated group and need only the untreated trend; the two assumption sets are non-nested, with ours weaker on trends but identifying a narrower estimand. Second, it conditions the trend assumption on the observed mediator, which is examinable within observed strata but exposed to collider bias, whereas we condition on the potential version $M(0)$. Third, its identification is for a binary mediator, with a continuous mediator handled only through a linear, additive regression, whereas ours holds for a general mediator. \citet{blackwell2025estimating} and \citet{huber2026difference} also condition on the realized mediator; the former targets controlled effects, which need no cross-world assumption but do not decompose the total effect. \citet{deuchert2019direct} and \citet{huber2022direct} are designed for experiments with non-compliance and a binary mediator and identify effects for compliers.

\begin{table}
\centering
\caption{Identification strategies for mediation analysis in difference-in-differences designs} \label{tab:compare}
\footnotesize
\setlength{\tabcolsep}{3pt}
\begin{tabular}{>{\raggedright\arraybackslash}p{2.5cm}*{6}{>{\raggedright\arraybackslash}p{2.05cm}}}
\toprule
 & \citet{deuchert2019direct} & \citet{huber2022direct} & \citet{blackwell2025estimating} & \citet{huber2026difference} & \citet{hsia2025causal} & This paper \\
\midrule
(i) Estimand; population & Natural effects; compliers & Natural effects; compliers & Controlled direct effects; population & Effects conditional on observed $M$, natural effects by integration; treated group & Natural effects; population & Natural effects; treated group (controlled effects in Section \ref{sec:control}) \\
(ii) Parallel trends required in & Untreated world, within principal strata & None (changes-in-changes) & Untreated world, within observed $M$ & Untreated world, within observed $M$ & Both treated and untreated worlds (homogeneous trends) & Untreated world only, within $M(0)$-strata \\
(iii) Unmeasured treatment--outcome confounding & Allowed (time-invariant) & Allowed & Allowed (time-invariant) & Allowed (time-invariant) & Allowed (time-invariant) & Allowed (time-invariant) \\
(iv) Cross-world assumption & Yes (natural effects) & Yes & No (controlled effects only) & For natural effects & Yes & Yes (Assumptions \ref{ass2}, \ref{ass3}(ii)) \\
(v) Key trend assumption examinable by pre-trends & Necessary condition only & Not applicable & Yes, within observed $M$ strata & Yes, within observed $M$ strata & Necessary condition only & Necessary condition only (Proposition \ref{prop:pt}) \\
(vi) Mediator & Binary & Binary & Discrete & General & Binary for identification; continuous via linear model & General \\
(vii) Functional form & Nonparametric & Nonparametric & No modified direct effect & Nonparametric & Linear, additive & Nonparametric \\
\bottomrule
\end{tabular}
\end{table}

In terms of guidance, a researcher who needs a population-level effect and accepts homogeneous trends and a linear mediator model is served by \citet{hsia2025causal}; one interested in controlled effects who accepts conditioning on the observed mediator in exchange for testable trends is served by \citet{blackwell2025estimating} or \citet{huber2026difference}; and one who wants natural effects in the treated group, is unwilling to assume homogeneous trends, has a non-binary mediator, or wishes to avoid conditioning on a post-treatment variable is served by the present framework. The identification-level gains of this paper are the weaker trend requirement and the general mediator; the estimation-level gains, developed next, are the efficient influence functions, multiple robustness, and the treatment of controlled effects for mixed mediators.

\section{Estimation} \label{sec:estimation}

\subsection{Efficient influence functions}

Theorem \ref{th:iden} suggests estimating $\tau(0,g^*)$ by imputation: regress $\Delta Y := Y_1-Y_0$ on $(M,X)$ in the control group to obtain $\widehat\delta(0,m,x)$, integrate it over an estimated mediator density $\widehat{f}(m \mid g^*,x)$ to obtain $\widehat\nu(g^*,X)$, and average over the treated group. This estimator is inconsistent if either $\delta(0,m,x)$ or $f(m\mid g^*,x)$ is misspecified, and it is inefficient.

To improve robustness and efficiency, we derive the efficient influence function (EIF) of $\tau(g,g^*)$ for $(g,g^*) \in \{(1,1), (0,0), (0,1)\}$. The EIF is the unique influence function of a regular and asymptotically linear (RAL) estimator that lies in the tangent space of the model, the set of score functions of all parametric submodels \citep{bickel1993efficient}; its variance is the nonparametric efficiency bound. Lemma 1 in the Supplementary Material characterizes the tangent space. Let $\pi(x) = P(G=1 \mid X=x)$ be the propensity score and $f(m \mid g,x)$ the conditional density (or probability) of the mediator.

\begin{theorem}\label{th:EIF}
Suppose Assumptions \ref{ass1}--\ref{ass5} hold. The efficient influence function of $\tau(1,1)$ is
\begin{align*}
\varphi(1,1) &= \frac{G}{P(G=1)} \{\Delta Y-\tau(1,1)\}.
\end{align*}
The efficient influence function of $\tau(0,0)$ is
\begin{align*}
\varphi(0,0) &= \frac{1-G}{P(G=1)} \frac{\pi(X)}{1-\pi(X)} \{\Delta Y - E(\delta(0,M,X)\mid G=0,X)\} \\
&\quad + \frac{G}{P(G=1)} \{E(\delta(0,M,X)\mid G=0,X) - \tau(0,0)\}.
\end{align*}
The efficient influence function of $\tau(0,1)$ is
\begin{align*}
\varphi(0,1) &= \frac{1-G}{P(G=1)} \frac{\pi(X)}{1-\pi(X)} \frac{f(M \mid G=1,X)}{f(M \mid G=0,X)} \{\Delta Y-\delta(0,M,X)\} \\
&\quad + \frac{G}{P(G=1)} \{\delta(0,M,X) - \tau(0,1)\}.
\end{align*}
The efficient influence functions of $\tau_{IE}$, $\tau_{DE}$, and $\tau$ are $\varphi_{IE} = \varphi(0,1)-\varphi(0,0)$, $\varphi_{DE} = \varphi(1,1)-\varphi(0,1)$, and $\varphi = \varphi(1,1)-\varphi(0,0)$, respectively.
\end{theorem}

The proof in the Supplementary Material verifies that $\varphi(g,g^*)$ lies in the tangent space and is a valid influence function for $\tau(g,g^*)$.

The efficient influence functions involve the propensity score, the outcome-change regression, and the conditional mediator density, the last entering in two places: the density ratio in $\varphi(0,1)$ and the integral $\nu(0,X) = E\{\delta(0,M,X) \mid G=0,X\}$ in $\varphi(0,0)$. Estimating a conditional density is difficult, and we avoid it in both places. For the ratio, Bayes' formula gives
\begin{equation}
\frac{\pi(X)}{1-\pi(X)} \frac{f(M \mid G=1,X)}{f(M \mid G=0,X)} = \frac{P(G=1 \mid M,X)}{P(G=0 \mid M,X)}, \label{bayes}
\end{equation}
so the ratio of propensity scores and mediator densities reduces to a pseudo-propensity score $\varpi(m,x) = P(G=1\mid M=m, X=x)$, which has a binary response and can be fitted by logistic regression, denoted $\widehat\varpi(m,x)$. Equation \eqref{bayes} holds whether $M$ is discrete, continuous, or mixed, provided the Radon--Nikodym derivative of the conditional distribution of $M$ in $G=1$ with respect to that in $G=0$ exists. 

For $\nu(0,x)$, two strategies avoid modeling $f(m \mid 0,x)$. The first is a plug-in, $\widehat\nu(0,x) = \widehat\delta\{0,\widehat{m}(x),x\}$ with $\widehat{m}(x)$ a regression estimate of $E(M \mid G=0,X=x)$. This is exact when $\delta(0,m,x) = a(x)+b(x)m$ is affine in $m$ for each $x$, which includes the working models used in the simulations and the application (linear in $M$ with $M \times X$ interactions), and it is the estimator we use. For specifications that are nonlinear in $m$, the plug-in is not valid; one instead regresses the pseudo-outcome $\widehat\delta(0,M_i,X_i)$ on $X_i$ in the control group and takes the fitted value as $\widehat\nu(0,x)$, with a second-stage basis rich enough to capture $E\{\delta(0,M,X) \mid G=0,X=x\}$.

\subsection{Estimators based on efficient influence functions}
For any measurable function $q(o)$, let $\mathbb{P}(q) = \int q(o)f(o)do$ and $\mathbb{P}_n(q) = n^{-1}\sum_{i=1}^{n} q(O_i)$, and let $\|q\|_{L_1} = \mathbb{P}(|q|)$ and $\|q\|_{L_2} = \{\mathbb{P}(q^2)\}^{1/2}$. Solving the empirical estimating equation $\mathbb{P}_n\widehat\varphi(g,g^*) = 0$ with fitted nuisance functions gives
\begin{align*}
\widehat\tau(1,1) &= \frac{1}{\mathbb{P}_n(G)} \mathbb{P}_n(G \Delta Y), \\
\widehat\tau(0,0) &= \frac{1}{\mathbb{P}_n(G)} \mathbb{P}_n \bigg[(1-G) \frac{\widehat\pi(X)}{1-\widehat\pi(X)} \{\Delta Y - \widehat\nu(0,X)\} + G \widehat\nu(0,X)\bigg], \\
\widehat\tau(0,1) &= \frac{1}{\mathbb{P}_n(G)} \mathbb{P}_n\bigg[(1-G) \frac{\widehat\varpi(M,X)}{1-\widehat\varpi(M,X)} \{\Delta Y -\widehat\delta(0,M,X)\} + G\widehat\delta(0,M,X)\bigg].
\end{align*}
The natural indirect, direct, and total effects are then estimated as
\[
\widehat\tau_{IE} = \widehat\tau(0,1)-\widehat\tau(0,0), \quad \widehat\tau_{DE} = \widehat\tau(1,1)-\widehat\tau(0,1), \quad \widehat\tau = \widehat\tau(1,1)-\widehat\tau(0,0).
\]
Next, we establish a general consistency result for the proposed estimators.

\begin{theorem}[Multiple robustness] \label{th:mr}
Suppose the working models $\{\pi(x), \varpi(m,x), \delta(0,m,x), \nu(0,x)\}$ are Glivenko--Cantelli. Then $\widehat\tau(1,1)$ is consistent for $\tau(1,1)$; $\widehat\tau(0,0)$ is consistent for $\tau(0,0)$ if either $\widehat\pi(x)$ or $\widehat\nu(0,x)$ is $L_1$-consistent; and $\widehat\tau(0,1)$ is consistent for $\tau(0,1)$ if either $\widehat\varpi(m,x)$ or $\widehat\delta(0,m,x)$ is $L_1$-consistent. Consequently, $\widehat\tau_{IE}$, $\widehat\tau_{DE}$, and $\widehat\tau$ are consistent if one of the following four pairs is $L_1$-consistent: $\{\widehat\pi, \widehat\varpi\}$, $\{\widehat\pi, \widehat\delta\}$, $\{\widehat\nu, \widehat\varpi\}$, or $\{\widehat\nu, \widehat\delta\}$.
\end{theorem}

The Glivenko--Cantelli condition requires that the working models are not too complex \citep{vaart2023empirical}; (generalized) linear and kernel regression models satisfy it. Theorem \ref{th:mr} is stated in terms of the four nuisance functions that enter the estimator. They can be derived from three primitive models, $\pi(x)$, $f(m \mid g,x)$, and $\delta(0,m,x)$, since $\varpi$ is a function of $(\pi,f)$ by \eqref{bayes} and $\nu(0,x) = \int \delta(0,m,x) f(m \mid 0,x) dm$. When derived compatibly, correct specification of any two of $\{\pi,f,\delta\}$ implies one of the four pairs is correct ($\{\pi,f\}\Rightarrow\{\pi,\varpi\}$, $\{f,\delta\}\Rightarrow\{\nu,\delta\}$), so the estimator is robust to misspecification of any one primitive. In practice, we model $\varpi$ and $\nu$ directly to avoid estimating $f(m\mid g,x)$; the fitted $\widehat\varpi$ and $\widehat\nu$ then need not be compatible with a common mediator model, and the applicable statement is the union of pairs in Theorem \ref{th:mr}.

\begin{theorem}[Nonparametric efficiency]
Suppose the working models $\{\pi(x), \varpi(m,x), \delta(0,m,x), \nu(0,x)\}$ are Donsker. Suppose that the fitted models $\widehat\pi(x)$, $\widehat\varpi(m,x)$, $\widehat\delta(0,m,x)$, and $\widehat\nu(0,x)$ are $L_2$-consistent and converge at a rate of $o_p(n^{-1/4})$. Then $\sqrt{n}\{\widehat\tau(g,g^*)-\tau(g,g^*)\} \xrightarrow{d} N\{0,E\varphi(g,g^*)^2\}$, and the asymptotic variance attains the nonparametric efficiency bound. As a result, $\sqrt{n}(\widehat\tau_{IE}-\tau_{IE}) \xrightarrow{d} N(0, E\varphi_{IE}^2)$, $\sqrt{n}(\widehat\tau_{DE}-\tau_{DE}) \xrightarrow{d} N(0, E\varphi_{DE}^2)$, $\sqrt{n}(\widehat\tau-\tau) \xrightarrow{d} N(0, E\varphi^2)$, and they are all nonparametrically efficient.
\end{theorem}

Standard parametric and semiparametric working models are Donsker, and the fitted models may converge more slowly than the parametric rate, for example under kernel smoothing. Complex machine learning methods such as random forests are not Donsker in general; the Donsker condition can then be removed by cross-fitting \citep{chernozhukov2018double}, at the cost of greater finite-sample variation in small samples. We use simple Donsker working models throughout. When some working models are misspecified but the conditions of Theorem \ref{th:mr} hold, $\widehat\tau(g,g^*)$ remains asymptotically linear, with an influence function equal to the EIF evaluated at the limiting models plus a remainder due to nuisance estimation; the asymptotic variance is then not the efficiency bound, but the bootstrap remains valid for inference.

Supplementary Material A outlines practical estimation using generalized linear models. Standard errors are computed from the fitted influence functions $\widehat\varphi(g,g^*)$:
\begin{align*}
s.e.(\widehat\tau_{IE}) &= \bigg[\frac{1}{n} \mathbb{P}_n\{\widehat\varphi(0,1)-\widehat\varphi(0,0)\}^2\bigg]^{1/2}, \quad 
s.e.(\widehat\tau_{DE}) = \bigg[\frac{1}{n} \mathbb{P}_n\{\widehat\varphi(1,1)-\widehat\varphi(0,1)\}^2\bigg]^{1/2}, \\
s.e.(\widehat\tau) &= \bigg[\frac{1}{n} \mathbb{P}_n\{\widehat\varphi(1,1)-\widehat\varphi(0,0)\}^2\bigg]^{1/2}.
\end{align*}
The $P$-values are calculated based on the normal approximation.

\section{Extension to controlled effects} \label{sec:control}

The controlled direct effect (CDE) is the effect of switching treatment while holding the mediator at a fixed value $m$,
\[
\tau_{DE}(m) = E\{Y_1(1,m)-Y_1(0,m) \mid G=1\}.
\]
With $\bar\tau(g,m) = E\{Y_1(g,m)-Y_0(0) \mid G=1\}$, $\tau_{DE}(m) = \bar\tau(1,m)-\bar\tau(0,m)$. Let $\boldsymbol{\delta}(\cdot)$ be the Dirac delta function.
\begin{theorem}\label{thm:EIF_for_control}
Under the above conditions, $\bar\tau(g,m)$ is identified as 
\[
\bar\tau(g,m) = E\{\delta(g,m,X) \mid G=1\}.
\]
Furthermore, the standard Gateaux derivative based on a point mass contamination path of $\bar\tau(g,m)$ is
\[
\begin{aligned}
\bar\varphi(g,m) & = \frac{I(G = g) \boldsymbol{\delta} (M - m)}{P(G = 1)} \frac{P(G = 1 \mid X)}{f(G = g, M = m \mid  X)} \big\{ \Delta Y - \delta (g, m, X)  \big\} \\
& ~~ ~~ + \frac{G}{P(G = 1)} \big\{\delta (g, m, X) -  \bar\tau (g, m)  \big\}.
\end{aligned}
\]
%     In particular,
%     \[
%         \begin{aligned}
%             \bar\varphi(0,m) &= \frac{1-G}{P(G=1)} \frac{\pi(X)}{1-\pi(X)} \frac{\boldsymbol{\delta}(M - m)}{f(M \mid G=0,X)} \{\Delta Y - \delta(0,M,X)\} \\
% &\qquad + \frac{G}{P(G=1)} \{\delta(0,m,X)-\bar\tau(0,m)\}, \\
% \bar\varphi(1,m) &= \frac{G}{P(G=1)} \frac{\boldsymbol{\delta}(M - m)}{f(M\mid G=1,X)} \{\Delta Y-\delta(1,M,X)\} \\
% &\qquad + \frac{G}{P(G=1)} \{\delta(1,m,X)-\bar\tau(1,m)\}.
%         \end{aligned}
%     \]
Specifically, when $M$ is discrete, the above Gateaux derivatives will be the efficient influence functions. 
\end{theorem}

When the mediator is discrete, the fitted influence function $\widehat\varphi(g,m)$ replaces the nuisance functions by their estimates. When the mediator is continuous, no efficient influence function exists, because the Dirac mass at $M=m$ does not belong to $L^2(\mathbb{P})$. We then replace the Dirac mass by a kernel: let $K$ be a kernel function and $K_h(x) = h^{-1}K(x/h)$ with bandwidth $h$, and define the smoothed influence function
\begin{align*}
\bar\varphi_h(g,m) & = \frac{I(G = g)K_h (M - m)}{P(G = 1)} \frac{P(G = 1 \mid X)}{f(G = g, M = m \mid  X)} \big\{ \Delta Y - \delta (g, m, X)  \big\} \\
&\quad + \frac{G}{P(G = 1)} \big\{\delta (g, m, X) -  \bar\tau (g, m)  \big\}.
\end{align*}
with fitted version $\widehat\varphi_h(g,m)$.
% \begin{align*}
% \widehat\varphi_h(0,m) &= \frac{1-G}{\mathbb{P}_n(G)} \frac{\widehat\pi(X)}{1-\widehat\pi(X)} \frac{K_h(M-m)}{\widehat f(M \mid G=0,X)} \{\Delta Y - \widehat\delta(0,M,X)\} \\
% &\qquad + \frac{G}{\mathbb{P}_n(G)} \{\widehat\delta(0,m,X)-\widehat \tau(0,m)\}, \\
% \widehat\varphi_h(1,m) &= \frac{G}{\mathbb{P}_n(G)} \frac{K_h(M-m)}{\widehat f(M \mid G=1,X)} \{\Delta Y - \widehat\delta(1,M,X)\} \\
% &\qquad + \frac{G}{\mathbb{P}_n(G)} \{\widehat \delta(1,m,X)-\widehat\tau(1,m)\}.
% \end{align*}
% \red{We drop the sub index $h$ when it does not cause confusion in estimation as well as in the parameter}.
Define the corresponding estimator as
\begin{align}\label{eq:controlled_est}
\widehat\tau(g, m) = \frac{\mathbb{P}_n \big[ G \widehat\delta(g, m, X) \big] }{\mathbb{P}_n(G)} + \mathbb{P}_n [\widehat\varphi(g,m)] \quad
\text{or} \quad   \widehat\tau_h(g, m) = \frac{\mathbb{P}_n \big[ G \widehat\delta(g, m, X) \big] }{\mathbb{P}_n(G)} + \mathbb{P}_n [\widehat\varphi_h(g,m)]
\end{align}
for discrete and continuous $M$, respectively. Supplementary Material A describes the implementation.

We use $f(m \mid g, x)$ for either the conditional probability or the conditional density of $M$, and in the continuous case assume $f(m \mid g, x)$ and $\delta(g, m, x)$ are twice continuously differentiable in $m$ near the target value. The following theorem gives the asymptotic properties of \eqref{eq:controlled_est} for scalar $M$.

\begin{theorem}\label{thm:control_asy}
Let the working models $\{\pi(x), \delta(g,m,x), f(m \mid g, x)\}$ and the fitted models $\widehat\pi(x), \widehat\delta(g,m,x), \widehat{f}(m \mid g, x)$ be $\mathcal{M}_{\text{working}}$ and $\mathcal{M}_{\text{fitted}}$, respectively.
\begin{itemize}
\item (Multiple robustness) Suppose $\mathcal{M}_{\text{working}}$ are Glivenko-Cantelli, and either $\widehat\delta$ or $(\widehat{f}, \widehat\pi)$ (for $g = 1$, $\widehat{f}$ is sufficient) in $\mathcal{M}_{\text{fitted}}$ are $L_1$-consistent. Then: (i) if $M$ is discrete, $\widehat\tau(g, m)$ is consistent; (ii) if $M$ is continuous,  then $\widehat\tau_h(g, m)$ is consistent for $\bar\tau(g, m)$ when $K$ is symmetric kernel with $h \to 0$ and $nh \to \infty$.
\item (Asymptotic normality) Suppose  $\mathcal{M}_{\text{working}}$ are Donsker and $\mathcal{M}_{\text{fitted}}$ are $L_2$-consistent and converge at a rate of $o_P(n^{- \kappa})$ with $\kappa \in (0, 1 / 4)$. Then: (i) if $M$ is discrete,  $\widehat\tau(g, m)$ is $\sqrt{n}$-consistent and achieves the nonparametric efficiency; (ii) if $M$ is continuous, {$\widehat\tau_h(g, m) - \texttt{bias}(g, m) h^2$ is $\sqrt{nh}$-consistent} when $K$ is symmetric kernel with bandwidth $h$ such that: $h \to 0$, $nh \to \infty$, $nh^5 \to 0$, and $\log (1 / h) n^{-4 \kappa} \to 0$. Here $\texttt{bias}(g, m)$ is the kernel bias term defined in the Supplementary Material.
\end{itemize}
\end{theorem}

Kernel smoothing slows the convergence rate of $\widehat\tau_h(g,m)$ below $n^{-1/2}$, although up to the scale $h$ the influence function remains Donsker over $(g,m)$. The extension to multivariate $M$ requires only modified bandwidth conditions. The estimator is doubly robust whether $M$ is discrete or continuous, and it attains the nonparametric efficiency bound when $M$ is discrete.

\paragraph{Mixed mediator distributions.} In our application the mediator, a proportion of weeks employed, has point masses at $0$ and $1$ and a continuous component on $(0,1)$. Neither case of Theorem \ref{thm:control_asy} applies to such a variable as a whole, and a kernel that smooths across an atom treats a mass point as part of a density. The remedy is to let $f(m \mid g,x)$ be the density with respect to the counting measure on the atoms plus the Lebesgue measure on the interior, and to apply Theorem \ref{thm:control_asy} componentwise. At an atom, $f(m \mid g,x) = P(M=m \mid G=g,X=x)$ and $\bar\tau(g,m)$ is estimated by the discrete estimator, which is $\sqrt{n}$-consistent and efficient. At an interior point, $f(m \mid g,x) = P(0<M<1 \mid G=g,X=x)\, f_{\mathrm{int}}(m \mid g,x)$ with $f_{\mathrm{int}}$ the conditional density given that $M$ is interior, and $\bar\tau(g,m)$ is estimated by the kernel-smoothed estimator on interior observations only, at the $\sqrt{nh}$ rate. Identification is unaffected since Theorem \ref{thm:EIF_for_control} holds pointwise in $m$. The interior smoothing bias matches Theorem \ref{thm:control_asy} with the interior sub-density; local-polynomial fitting mitigates boundary bias near the atoms, where the effective sample size shrinks. Variances come from the (smoothed) influence function of each component. Unlike the natural effects, where the mediator density enters only through a ratio that reduces to the pseudo-propensity score \eqref{bayes}, the controlled effect at an interior point requires the \emph{level} $f(m \mid g,x)$, and this term cannot be removed without losing double robustness; at the atoms, and for a categorized mediator, only conditional probabilities are needed. In principle, explicit density estimation can be avoided by learning the Riesz representer of the smoothed functional, which is precisely the kernel weight in $\bar\varphi_h(g,m)$, directly from an empirical loss, combined with cross-fitting \citep{chernozhukov2022automatic}; we leave this extension to future work. Supplementary Material A describes the implementation.

% The bandwidth $h \to 0$ as $n \to \infty$. Since $E\{\bar\varphi(g,m)\} = 0$, the estimator $\widehat{\bar\tau}(g,m)$ obtained by solving the estimating equation is consistent as long as $\widehat\delta(g,m,x)$, the working model for $\delta(g,m,x)$ is consistent. If $\delta(g,m,x)$ is misspecified but other working models are consistent, then the bias is $O(h^2)$.
% Under smoothness conditions for $\widehat\delta(g,m,x)-\delta(g,m,x)$, the reminder term is $O(h^2)$, a higher-order term compared with $\mathbb{P}_n\bar\varphi(g,m)^2 = O(h^{-1})$ by choosing $h = o(n^{-1/3})$. Thus, $\widehat{\bar\tau}(g,m)$ converges at a $\sqrt{nh}$ rate. 

\section{Simulation} \label{sec:simulation}

In this section, we conduct simulation studies to compare our proposed method with the regression-based method in \citet{hsia2025causal}, which uses regression coefficients to represent treatment effects as an extension to \citet{baron1986moderator}. For reproducibility, the regression-based method is implemented as follows: we fit the outcome-change regression $E(Y_1-Y_0 \mid G,M,X) = \alpha_0 + \alpha_G G + \alpha_M M + \alpha_X^{\top}X$ and the mediator regression $E(M \mid G,X) = \gamma_0 + \gamma_G G + \gamma_X^{\top}X$ by ordinary least squares, and estimate the natural direct effect by $\widehat\alpha_G$, the natural indirect effect by $\widehat\alpha_M\widehat\gamma_G$, and the total effect by their sum. No interaction terms (neither $G \times M$ nor $M \times X$) are included, consistent with the linear, additive specification of \citet{hsia2025causal}. Standard errors are obtained from the influence functions of the least-squares coefficients. 

Let the sample size $n \in \{200, 1000, 5000\}$. We generate two independent baseline covariates $X_i = (X_{1i},X_{2i})^{\top}$, with each element following the standard normal distribution. In addition, we generate an individual-level random effect $u_i \sim N(0,1)$. We consider two settings for generating the mediator. In Setting 1, the potential mediator is continuous, generated as
\[
M_i(g) = 0.6X_{1i} - 0.3X_{2i} + g + \varepsilon_i,
\]
where $\varepsilon_i \sim N(0,1)$. In Setting 2, the potential mediator is binary, generated from
\[
P(M_i(g)=1 \mid X_0) = \Phi(0.6X_{1i} - 0.3X_{2i} + g),
\]
where $\Phi(\cdot)$ is the cumulative distribution function of the standard normal distribution. The potential outcomes are generated as
\begin{align*}
Y_{0i}(0) &= 2X_{1i} + u_i + \epsilon_{0i}, \\
Y_{1i}(g,m) &= X_{1i} + X_{2i} + g + 0.5(1+0.4X_{2i})m + u_i + \epsilon_{1i},
\end{align*}
where $\epsilon_{0i}$ and $\epsilon_{1i}$ are independent random errors following $N(0,0.5^2)$.
Finally, the treatment assignment mechanism is
\[
P(G_i=1 \mid X_i) = \frac{\exp(0.3+0.4X_1+0.5X_2)}{1+\exp(0.3+0.4X_1+0.5X_2)}.
\]
The observed mediator and outcomes $M_i=M_i(G_i)$, $Y_{0i}=Y_{0i}(0)$, and $Y_{1i}=Y_{1i}(G_i,M_i)$.

For the proposed method, the propensity score is modeled by logistic regression, and the change in observed outcomes is modeled by linear regression with interactions between $M$ and $X$ in the control group, both of which are correctly specified. The pseudo-propensity score $\varpi(m,x)$ is modeled by logistic regression with $M \times X$ interactions following \eqref{bayes}, and $\nu(0,x)$ by the plug-in described in Section \ref{sec:estimation}, which is exact here because $\delta(0,m,x)$ is affine in $m$. We replicate the data generation 1000 times. The Panel (O) of Table \ref{tab1} shows the average bias, average standard error (SE), standard deviation (SD), and coverage percentage (CP) of 95\% confidence intervals for the estimated natural indirect, direct, and total effects. The true effects in the treated group are computed by Monte Carlo from the potential outcomes with $2\times10^6$ units. We observe that the regression-based method exhibits considerable bias in estimating $\tau_{IE}$ and $\tau$, because the linear model does not account for the interaction between $M$ and $X$, and its coverage deteriorates as $n$ grows. In contrast, the bias of the proposed method is negligible, the average standard error and standard deviation are in close agreement, and the coverage percentage is close to the nominal level for $n \geq 1000$.

\begin{table}
\centering
\caption{Simulation results for the estimated natural indirect, direct, and total effects} \label{tab1}
\setlength{\tabcolsep}{4pt}
\begin{tabular}{llccccccccccccc}
\toprule
 & & \multicolumn{6}{c}{Setting 1: continuous mediator} & & \multicolumn{6}{c}{Setting 2: binary mediator} \\
 \cmidrule{3-8}  \cmidrule{10-15}
$n$ & & \multicolumn{3}{c}{Regression-based} & \multicolumn{3}{c}{Proposed} & & \multicolumn{3}{c}{Regression-based} & \multicolumn{3}{c}{Proposed} \\
 \cmidrule{3-5} \cmidrule{6-8}  \cmidrule{10-12} \cmidrule{13-15}
 & & $\tau_{IE}$ & $\tau_{DE}$ & $\tau$ & $\tau_{IE}$ & $\tau_{DE}$ & $\tau$ & & $\tau_{IE}$ & $\tau_{DE}$ & $\tau$ & $\tau_{IE}$ & $\tau_{DE}$ & $\tau$ \\
\midrule
\multicolumn{15}{l}{(O) Models correctly specified} \\
200 & Bias & -.039 & -.009 & -.048 & .002 & .007 & .009 && -.009 & -.004 & -.013 & .001 & -.004 & -.002 \\
 & SE & .094 & .122 & .133 & .143 & .145 & .151 && .050 & .112 & .111 & .079 & .127 & .122 \\
 & SD & .094 & .126 & .135 & .174 & .180 & .161 && .052 & .116 & .115 & .083 & .138 & .131 \\
 & CP & .903 & .947 & .936 & .892 & .883 & .932 && .911 & .946 & .945 & .923 & .928 & .931 \\
1000 & Bias & -.041 & -.012 & -.054 & -.002 & .001 & -.001 && -.011 & -.001 & -.012 & -.001 & .002 & .001 \\
 & SE & .042 & .055 & .060 & .069 & .071 & .068 && .022 & .050 & .050 & .034 & .057 & .054 \\
 & SD & .042 & .057 & .062 & .075 & .077 & .071 && .022 & .051 & .050 & .034 & .057 & .055 \\
 & CP & .806 & .938 & .853 & .942 & .935 & .936 && .896 & .938 & .937 & .941 & .946 & .945 \\
5000 & Bias & -.038 & -.014 & -.052 & .001 & -.001 & .001 && -.010 & -.002 & -.013 & -.000 & .001 & .001 \\
 & SE & .019 & .025 & .027 & .031 & .033 & .031 && .010 & .023 & .022 & .015 & .026 & .024 \\
 & SD & .018 & .025 & .027 & .033 & .033 & .030 && .010 & .023 & .023 & .015 & .026 & .025 \\
 & CP & .470 & .912 & .503 & .928 & .957 & .951 && .796 & .949 & .903 & .945 & .954 & .954 \\
\midrule
\multicolumn{15}{l}{(A) Outcome model misspecified} \\
200 & Bias & .031 & -.119 & -.089 & -.067 & .125 & .058 && .053 & -.129 & -.076 & .004 & .048 & .052 \\
 & SE & .162 & .303 & .276 & .363 & .540 & .436 && .117 & .282 & .263 & .262 & .486 & .420 \\
 & SD & .171 & .302 & .276 & .537 & .704 & .424 && .121 & .279 & .259 & .323 & .518 & .411 \\
 & CP & .940 & .933 & .943 & .875 & .877 & .957 && .948 & .918 & .938 & .965 & .929 & .957 \\
1000 & Bias & .028 & -.135 & -.107 & -.013 & .017 & .003 && .038 & -.132 & -.094 & .005 & -.005 & .000 \\
 & SE & .073 & .136 & .124 & .206 & .300 & .209 && .051 & .129 & .120 & .118 & .234 & .204 \\
 & SD & .074 & .140 & .127 & .258 & .336 & .189 && .050 & .129 & .121 & .119 & .219 & .179 \\
 & CP & .947 & .822 & .846 & .899 & .921 & .965 && .907 & .818 & .878 & .956 & .954 & .962 \\
5000 & Bias & .030 & -.136 & -.106 & -.002 & .008 & .006 && .042 & -.132 & -.089 & .012 & -.014 & -.001 \\
 & SE & .033 & .061 & .055 & .104 & .145 & .095 && .023 & .058 & .054 & .055 & .109 & .093 \\
 & SD & .032 & .060 & .054 & .115 & .140 & .082 && .022 & .059 & .055 & .053 & .100 & .083 \\
 & CP & .856 & .394 & .517 & .924 & .940 & .981 && .573 & .369 & .612 & .963 & .972 & .971 \\
\midrule
\multicolumn{15}{l}{(B) Propensity score misspecified} \\
200 & Bias & -.050 & .012 & -.038 & .006 & -.007 & -.001 && -.014 & .010 & -.004 & .000 & .001 & .002 \\
 & SE & .096 & .128 & .138 & .153 & .157 & .154 && .051 & .118 & .116 & .099 & .144 & .124 \\
 & SD & .099 & .128 & .141 & .209 & .214 & .188 && .053 & .121 & .117 & .117 & .177 & .147 \\
 & CP & .864 & .957 & .941 & .844 & .840 & .890 && .891 & .939 & .951 & .911 & .904 & .904 \\
1000 & Bias & -.052 & .019 & -.033 & -.002 & -.001 & -.003 && -.014 & .010 & -.004 & -.002 & .004 & .002 \\
 & SE & .043 & .057 & .062 & .079 & .082 & .070 && .023 & .053 & .052 & .044 & .066 & .055 \\
 & SD & .042 & .057 & .063 & .092 & .097 & .081 && .024 & .053 & .052 & .046 & .072 & .063 \\
 & CP & .769 & .940 & .907 & .907 & .911 & .902 && .859 & .948 & .953 & .929 & .919 & .909 \\
5000 & Bias & -.051 & .019 & -.032 & .002 & -.002 & -.000 && -.013 & .009 & -.005 & -.000 & -.000 & -.000 \\
 & SE & .019 & .026 & .028 & .038 & .040 & .031 && .010 & .024 & .023 & .020 & .030 & .025 \\
 & SD & .019 & .026 & .028 & .043 & .044 & .036 && .010 & .024 & .023 & .020 & .031 & .028 \\
 & CP & .251 & .872 & .783 & .936 & .930 & .918 && .722 & .942 & .941 & .941 & .937 & .907 \\
\bottomrule
\end{tabular}
\end{table}

To demonstrate the potential bias in estimation and inference when models are misspecified, we consider two alternative data-generating mechanisms. First, we assume the potential outcomes are generated as
\begin{align*}
Y_{0i}(0) &= 2X_{1i}\log(1+|X_{2i}|) + u_i + \epsilon_{0i}, \\
Y_{1i}(g,m) &= (X_{1i}+X_{2i})X_{2i} + g + 0.5(1+0.5gX_{2i})m + u_i + \epsilon_{1i},
\end{align*}
where $\epsilon_{0i}$ and $\epsilon_{1i}$ are independent random errors following $N(0,0.5^2)$, so that the outcome model is misspecified. Second, we assume the treatment assignment is generated from
\[
P(G_{i}=1 \mid X_i) = \Phi(0.3+0.4X_{1i}+0.5X_{2i}+0.3X_{1i}X_{2i}),
\]
so that the propensity score is misspecified. 

The Panels (A) and (B) of Table \ref{tab1} show the average bias, average standard error (SE), standard deviation (SD), and coverage percentage (CP) of 95\% confidence intervals for the estimated natural indirect, direct, and total effects when either the outcome model or propensity score is misspecified. The regression-based method shows considerable bias because the linear model does not correctly specify the data-generating mechanism. Even if a working model is misspecified, the bias of the proposed method remains small, and the coverage percentage of the 95\% confidence intervals is near the nominal level, as implied by Theorem \ref{th:mr}: in Panel (A) consistency rests on $\{\pi,\varpi\}$ and in Panel (B) on $\{\nu,\delta\}$. In these cases the proposed method is not efficient, and its standard errors are larger than under correct specification.

Finally, we compare the proposed estimator of the controlled effects $\bar\tau(1,0)$, $\bar\tau(0,0)$, and $\tau_{DE}(0)$ in the treated group with a regression-based benchmark that fits $E(Y_1-Y_0 \mid G,M,X)$ by linear regression with a $G \times M$ interaction, imputes the outcome change at $m=0$, and averages over the treated group. For the binary mediator, $m=0$ is an atom and we use the discrete estimator of Theorem \ref{thm:control_asy} with $P(M=0 \mid G=g,X)$ estimated by probit regression. For the continuous mediator, we use the kernel-smoothed estimator with a Gaussian kernel, a local-quadratic fit in $m$ that is linear in $X$, the conditional density $f(m \mid g,x)$ estimated by a normal linear model within each group, and bandwidth $h = 0.9\min\{\mathrm{sd}(M), \mathrm{IQR}(M)/1.34\}\, n^{-0.3}$, i.e., Silverman's constant \citep{silverman2018density} with an undersmoothing exponent so that $nh^5 \to 0$ as Theorem \ref{thm:control_asy} requires; with the $n^{-1/5}$ exponent the $h^2$ bias term is of the same order as the standard error and coverage drops to about 0.91. Panels (A) and (B) misspecify the outcome regression and the propensity score as in Table \ref{tab1}. Table \ref{tab2} reports the results. The regression-based benchmark is biased even under correct specification because it omits the $M \times X$ interaction. The proposed estimator is nearly unbiased in all three panels for $n \geq 1000$, with coverage between 0.90 and 0.96 at $n=5000$; its standard errors are larger for the continuous mediator, reflecting the $\sqrt{nh}$ rate, and it undercovers at $n=200$, where the effective sample size near $m=0$ is small.

\begin{table}
\centering
\caption{Simulation results for the estimated controlled effects $\bar\tau(1,0)$, $\bar\tau(0,0)$, and $\tau_{DE}(0)$ in the treated group} \label{tab2}
\setlength{\tabcolsep}{1pt}
\begin{tabular}{llccccccccccccc}
\toprule
& & \multicolumn{6}{c}{Setting 1: continuous mediator} & & \multicolumn{6}{c}{Setting 2: binary mediator} \\
 \cmidrule{3-8}  \cmidrule{10-15}
$n$ & & \multicolumn{3}{c}{Regression-based} & \multicolumn{3}{c}{Proposed} & & \multicolumn{3}{c}{Regression-based} & \multicolumn{3}{c}{Proposed} \\
 \cmidrule{3-5} \cmidrule{6-8}  \cmidrule{10-12} \cmidrule{13-15}
& & $\bar\tau(1,0)$ & $\bar\tau(0,0)$ & $\tau_{DE}(0)$ & $\bar\tau(1,0)$ & $\bar\tau(0,0)$ & $\tau_{DE}(0)$ & & $\bar\tau(1,0)$ & $\bar\tau(0,0)$ & $\tau_{DE}(0)$ & $\bar\tau(1,0)$ & $\bar\tau(0,0)$ & $\tau_{DE}(0)$ \\
\midrule
\multicolumn{15}{l}{(O) Models correctly specified} \\
200 & Bias & -.047 & -.014 & -.033 & .019 & -.009 & .028 && -.048 & .034 & -.082 & .004 & -.004 & .009 \\
 & SE & .099 & .089 & .132 & .211 & .208 & .232 && .160 & .116 & .191 & .221 & .197 & .232 \\
 & SD & .163 & .164 & .131 & .235 & .240 & .281 && .206 & .186 & .196 & .297 & .219 & .319 \\
 & CP & .744 & .721 & .956 & .916 & .893 & .884 && .866 & .776 & .921 & .847 & .923 & .818 \\
1000 & Bias & -.038 & -.004 & -.034 & .015 & -.000 & .015 && -.049 & .033 & -.083 & .002 & -.002 & .004 \\
 & SE & .044 & .039 & .058 & .118 & .116 & .145 && .071 & .052 & .085 & .120 & .093 & .128 \\
 & SD & .071 & .069 & .057 & .136 & .124 & .168 && .094 & .084 & .087 & .131 & .096 & .138 \\
 & CP & .708 & .729 & .916 & .916 & .941 & .913 && .798 & .730 & .846 & .941 & .940 & .927 \\
5000 & Bias & -.041 & -.007 & -.034 & .008 & -.000 & .008 && -.049 & .033 & -.082 & .002 & -.002 & .004 \\
 & SE & .020 & .018 & .026 & .066 & .064 & .085 && .032 & .023 & .038 & .058 & .043 & .062 \\
 & SD & .032 & .034 & .026 & .071 & .067 & .092 && .042 & .037 & .038 & .065 & .044 & .067 \\
 & CP & .457 & .677 & .734 & .935 & .946 & .930 && .627 & .628 & .417 & .931 & .959 & .938 \\
\midrule
\multicolumn{15}{l}{(A) Outcome model misspecified} \\
200 & Bias & -.004 & .115 & -.119 & .002 & -.186 & .188 && -.051 & .020 & -.071 & .018 & -.024 & .042 \\
 & SE & .264 & .236 & .350 & .375 & .544 & .691 && .448 & .325 & .535 & .433 & .522 & .725 \\
 & SD & .261 & .236 & .328 & .536 & .784 & .974 && .443 & .370 & .558 & .722 & .656 & .941 \\
 & CP & .949 & .932 & .955 & .852 & .788 & .836 && .959 & .905 & .933 & .828 & .893 & .899 \\
1000 & Bias & -.009 & .109 & -.118 & -.003 & -.025 & .022 && -.044 & .005 & -.048 & .010 & .003 & .006 \\
 & SE & .118 & .105 & .156 & .238 & .399 & .479 && .198 & .145 & .237 & .254 & .266 & .384 \\
 & SD & .112 & .103 & .140 & .298 & .477 & .567 && .199 & .157 & .236 & .334 & .281 & .453 \\
 & CP & .956 & .837 & .904 & .909 & .873 & .910 && .946 & .915 & .942 & .930 & .952 & .941 \\
5000 & Bias & -.016 & .105 & -.121 & -.001 & -.014 & .013 && -.048 & .007 & -.055 & .001 & -.002 & .002 \\
 & SE & .053 & .047 & .070 & .154 & .248 & .301 && .089 & .065 & .106 & .128 & .136 & .192 \\
 & SD & .051 & .047 & .065 & .192 & .272 & .330 && .087 & .069 & .100 & .154 & .136 & .199 \\
 & CP & .950 & .379 & .600 & .936 & .918 & .925 && .924 & .923 & .940 & .949 & .975 & .965 \\
\midrule
\multicolumn{15}{l}{(B) Propensity score misspecified} \\
200 & Bias & -.029 & -.019 & -.010 & .025 & -.015 & .040 && -.037 & .054 & -.091 & .005 & .008 & -.003 \\
 & SE & .098 & .096 & .135 & .198 & .206 & .235 && .159 & .123 & .193 & .220 & .190 & .240 \\
 & SD & .153 & .158 & .136 & .229 & .272 & .323 && .202 & .179 & .199 & .298 & .233 & .340 \\
 & CP & .785 & .764 & .939 & .891 & .834 & .818 && .875 & .807 & .912 & .851 & .891 & .834 \\
1000 & Bias & -.022 & -.017 & -.005 & .010 & -.012 & .022 && -.035 & .042 & -.077 & -.000 & -.006 & .006 \\
 & SE & .043 & .042 & .060 & .114 & .118 & .148 && .070 & .054 & .085 & .114 & .092 & .127 \\
 & SD & .068 & .071 & .058 & .125 & .151 & .185 && .090 & .080 & .086 & .129 & .109 & .154 \\
 & CP & .776 & .757 & .955 & .919 & .878 & .889 && .842 & .755 & .853 & .932 & .909 & .917 \\
5000 & Bias & -.022 & -.015 & -.007 & .007 & .002 & .004 && -.037 & .046 & -.083 & -.003 & -.002 & -.001 \\
 & SE & .019 & .019 & .027 & .065 & .067 & .088 && .031 & .024 & .038 & .054 & .042 & .060 \\
 & SD & .031 & .032 & .026 & .068 & .076 & .096 && .038 & .036 & .038 & .055 & .048 & .065 \\
 & CP & .683 & .710 & .953 & .936 & .914 & .921 && .733 & .506 & .403 & .958 & .902 & .929 \\
\bottomrule
\end{tabular}
\end{table}

\section{Application to Job Corps Study} \label{sec:application}

\subsection{Data}

The U.S. Job Corps experimental study is a job training program for disadvantaged youths \citep{schochet2001national}. Participants in the treated group are provided access to academic and vocational training. Approximately 60.4\% of the 9240 participants were assigned to the treated group. Although the treatment assignment was randomized, not all participants complied with the assignment. In the first year after assignment, only 71.1\% actually enrolled in education and/or vocational training. As a result, the realized treatment is no longer randomized; in particular, individuals with more pessimistic expectations about their labor market prospects were more likely to take up training. Participants spent substantial time in training, leaving less time for work, so earnings tend to be lower when they cannot work full-time; existing studies did not consider this and reckon that job training has only long-term effects \citep{schochet2008does}. We adjust for the treatment effect on earnings mediated by work time, using data from the R package \texttt{causalweight}. The treatment $G$ is enrollment in education and/or vocational training during the first year after assignment. The pre-treatment outcome $Y_0$ is average weekly gross earnings at assignment, which precedes any training. We consider two post-treatment outcomes $Y_1$, average weekly earnings in the second (short run) and third (long run) year after assignment, with the mediator $M$ the proportion of weeks employed in the same year, rescaled to $[0,1]$. Both analyses share the treatment and pre-treatment outcome and differ only in the horizon, so the pre/post ordering is unambiguous.

Weekly earnings are averaged over all weeks of the year, so they are zero for individuals who did not work (87.1\% at assignment, 24.8\% in year 2, 17.3\% in year 3); in the data, earnings are zero if and only if the proportion of weeks employed is zero (seven exceptions). We analyze $\log(1+\text{earnings})$. Because treatment can affect whether earnings are positive, differences in $\log(1+Y)$ have no percentage interpretation and depend on the units of measurement \citep{chen2024logs}; we therefore describe all estimates as effects on transformed earnings. Mean positive earnings (151 at assignment, 175 in year 2, 209 in year 3) are far above one, so the constant is immaterial for the employed, but this does not resolve the extensive-margin issue.

The main specification conditions on all baseline covariates measured at assignment: sex, age, race (white, black, Hispanic), years of education, GED and high school degrees, English mother tongue, cohabitation or marriage, having children, having ever worked, household size, parents' years of education, childhood welfare receipt, general health, smoking, alcohol consumption, and missingness indicators (25 covariates; three nearly identical missingness indicators are merged). This set covers prior labor-market attachment, health, and family background, which could plausibly affect training participation, employment intensity, and earnings growth jointly. %As a robustness check, we also report the four covariates (sex, age, education, black) most often used in previous work with these data. The pre-treatment outcome enters through the trend, not through $X$.

\subsection{Identifying assumptions}

%In year 2, 23.7\% of the control group ($G=0$) and 25.1\% of the treated group ($G=1$) did not work, 19.8\% and 14.8\% worked every week, and the average proportion of weeks employed is 0.490 and 0.449. In year 3 the corresponding figures are 18.6\% and 16.7\%, 25.1\% and 22.7\%, and 0.550 and 0.556. The average change in $\log(1+\text{earnings})$ from assignment is 3.07 (control) and 3.00 (treated) to year 2, and 3.46 and 3.63 to year 3; the large changes reflect that most participants had zero earnings at assignment.

No anticipation requires that earnings at assignment are unaffected by subsequent enrollment. In the data, pre-treatment earnings are measured at assignment, which precedes enrollment in a center, so it is natural to assume the enrollment does not affect pre-treatment earnings. Consistent with this, baseline earnings are balanced across eventual enrollers and non-enrollers (means 18.2 versus 22.3, covariate-adjusted $P=0.81$; shares with positive earnings 12.6\% versus 13.6\%, $P=0.32$), although with a single pre-treatment period this is balance in levels rather than the pre-trend test of Proposition \ref{prop:pt}.

Mediator-adjusted parallel trends requires that, among individuals who would have the same employment intensity without training, the change in transformed earnings in the absence of training would be the same for enrollers and non-enrollers given covariates. Enrollers may thus differ in persistent ways that shift earnings levels (motivation, ability, pessimism about job prospects) but not earnings growth. Sequential ignorability requires that, given covariates, enrollment is unconfounded with potential employment intensities and that employment intensity does not modify the untreated trend. The framework excludes treatment-induced confounders, such as training-induced changes in job search, and they are untestable. Positivity is supported: estimated propensity scores range from 0.29 to 0.91 and pseudo-propensity scores from 0.15 to 0.93.

\subsection{Natural effects}

Table \ref{tab4} reports the estimates. In the main specification (continuous mediator), the total effect on year-2 transformed earnings is 0.070 (s.e.\ 0.063, 95\% CI $[-0.054, 0.193]$); the natural indirect effect is $-0.035$ (s.e.\ 0.045), consistent with training slightly reducing earnings through reduced employment, though not significantly; and the natural direct effect is 0.105 (s.e.\ 0.046, 95\% CI $[0.015, 0.194]$, $P=0.022$). Holding employment intensity at the level realized under training, enrollment increases year-2 earnings; the insignificant short-run total effect is the net of a positive direct effect and a negative employment channel.

In year 3, the total effect is 0.283 (s.e.\ 0.061, 95\% CI $[0.163, 0.402]$), the natural indirect effect 0.191 (s.e.\ 0.044, 95\% CI $[0.105, 0.278]$), and the natural direct effect 0.091 (s.e.\ 0.044, 95\% CI $[0.004, 0.178]$, $P=0.040$). Two-thirds of the long-run effect operates through higher employment intensity, consistent with training first displacing work and then improving employment prospects, and one-third is direct. This aligns with the long-run gains found by instrumental-variable analyses \citep{schochet2008does}, but attributes most of the gain to the employment channel.

The proportion of weeks employed is a component of annual earnings: weekly earnings averaged over the year equal earnings per employed week times the proportion of weeks employed, so $Y_1 = 0$ whenever $M=0$, and part of the estimated indirect effect arises from the construction of the outcome rather than from a separate economic mechanism. Two consequences follow. First, $Y_1(g,m)$ is the transformed weekly earnings that would be observed under exposure $g$ if employment intensity were externally fixed at $m$, for instance through the availability of work; for $m=0$ it is zero regardless of $g$, so the controlled direct effect at $m=0$ is zero by construction. Second, a degenerate $Y_1(g,0)\equiv 0$ means that within the $M(0)=0$ stratum the ``trend'' $Y_1(0,0)-Y_0(0)$ equals $-Y_0(0)$, so Assumption \ref{ass2} there requires equal mean baseline earnings across groups, a restriction on levels rather than trends. This interaction between the mechanical link and the difference-in-differences logic is a reason to prefer an outcome that is not degenerate at $m=0$.

The remaining rows of Table \ref{tab4} assess robustness. Categorizing the mediator into four levels, $0$ ($M=0$), $1$ ($0<M\leq0.5$), $2$ ($0.5<M<1$), and $3$ ($M=1$), and applying the discrete estimator leaves signs and magnitudes unchanged. Conditioning only on the four covariates used in earlier analyses of these data (sex, age, education, black) gives a year-2 direct effect of 0.106 (s.e.\ 0.045) and year-3 indirect, direct, and total effects of 0.170, 0.092, and 0.262 (s.e.\ 0.043, 0.043, 0.060), close to the main specification. Estimating $\nu(0,x)$ by a linear second-stage regression instead of the (exact) plug-in changes the estimates by at most 0.004 (second block of Table \ref{tab4}); the direct effect is unaffected because it does not involve $\nu$. The last row of Table \ref{tab4} uses as outcome $\log(1+\text{earnings per employed week})$, with earnings per employed week equal to weekly earnings divided by the proportion of weeks employed for those who worked and zero otherwise; earnings at assignment are already a weekly rate, so $Y_0$ is unchanged. This outcome is a wage-like intensity measure for the employed, no longer proportional to $M$, though it retains an extensive margin. The estimates are close to the main specification: in year 2 the direct effect is 0.114 (s.e.\ 0.057, $P=0.044$) and the indirect effect $-0.025$ (s.e.\ 0.039); in year 3 the total effect is 0.248 (s.e.\ 0.064), the indirect effect 0.158 (s.e.\ 0.038), and the direct effect 0.091 (s.e.\ 0.053, $P=0.083$). The long-run indirect effect thus survives removal of the mechanical component, indicating that training raises the earnings rate partly by moving participants into more stable employment. Hourly wages are not in the public-use extract.

\begin{table}
\centering
\caption{Estimated natural indirect, direct, and total effects of year-1 training on $\log(1+\text{weekly earnings})$ in the Job Corps Study. Standard errors in parentheses. $^{*}P<0.05$, $^{**}P<0.01$, $^{***}P<0.001$.} \label{tab4}
\begin{tabular}{lcccccc}
\toprule
  Specification & \multicolumn{3}{c}{Year 2 outcome} & \multicolumn{3}{c}{Year 3 outcome} \\
  \cmidrule(lr){2-4} \cmidrule(lr){5-7}
  & $\tau_{IE}$ & $\tau_{DE}$ & $\tau$ & $\tau_{IE}$ & $\tau_{DE}$ & $\tau$ \\
  \midrule
  \multicolumn{7}{l}{\emph{Main}} \\
  Continuous $M$ & $-$0.0351 & 0.1046 & 0.0695 & 0.1914 & 0.0912 & 0.2825 \\
  & (0.0447) & (0.0455)$^{*}$ & (0.0630) & (0.0442)$^{***}$ & (0.0443)$^{*}$ & (0.0612)$^{***}$ \\
  \multicolumn{7}{l}{\emph{Robustness: $\nu(0,x)$ by second-stage regression}} \\
  Continuous $M$ & $-$0.0385 & 0.1046 & 0.0661 & 0.1892 & 0.0912 & 0.2804 \\
  & (0.0447) & (0.0455)$^{*}$ & (0.0630) & (0.0442)$^{***}$ & (0.0443)$^{*}$ & (0.0612)$^{***}$ \\
  \multicolumn{7}{l}{\emph{Robustness: ordinal $M$}} \\
  Ordinal $M$ & $-$0.0578 & 0.1294 & 0.0716 & 0.1679 & 0.1148 & 0.2828 \\
  & (0.0451) & (0.0449)$^{**}$ & (0.0630) & (0.0446)$^{***}$ & (0.0442)$^{**}$ & (0.0612)$^{***}$ \\
  \multicolumn{7}{l}{\emph{Robustness: outcome $=\log(1+\text{earnings per employed week})$}} \\
  Continuous $M$ & $-$0.0253 & 0.1136 & 0.0883 & 0.1575 & 0.0908 & 0.2484 \\
  & (0.0388) & (0.0565)$^{*}$ & (0.0671) & (0.0380)$^{***}$ & (0.0525) & (0.0641)$^{***}$ \\
  \bottomrule
\end{tabular}
\end{table}

\subsection{Controlled direct effects}

We estimate the controlled direct effect (CDE) $\tau_{DE}(m)$, the effect of training on transformed earnings growth had employment intensity been fixed at $m$, with the componentwise estimator of Section \ref{sec:control}: at the atoms $m=0$ and $m=1$, the discrete estimator with $P(M=m \mid G=g,X)$ from multinomial logistic regression; at $m \in \{0.05, \ldots, 0.95\}$, the kernel-smoothed estimator on interior observations with a Gaussian kernel, Silverman's rule-of-thumb bandwidth from the interior observations ($h \approx 0.045$; employment is recorded in weeks, so $M$ lies on a grid with spacing $0.019$ and the undersmoothed rate of Section \ref{sec:simulation} would fall below it), a local-quadratic fit in $m$ linear in $X$, and the interior density modeled as normal on the logit scale. The effective sample size of the kernel weights exceeds 100 (control) and 250 (treated) at every interior grid point. Standard errors come from the (smoothed) efficient influence functions. We also report the ordinal specification.

\begin{figure}
    \centering
    \includegraphics[width=0.95\textwidth]{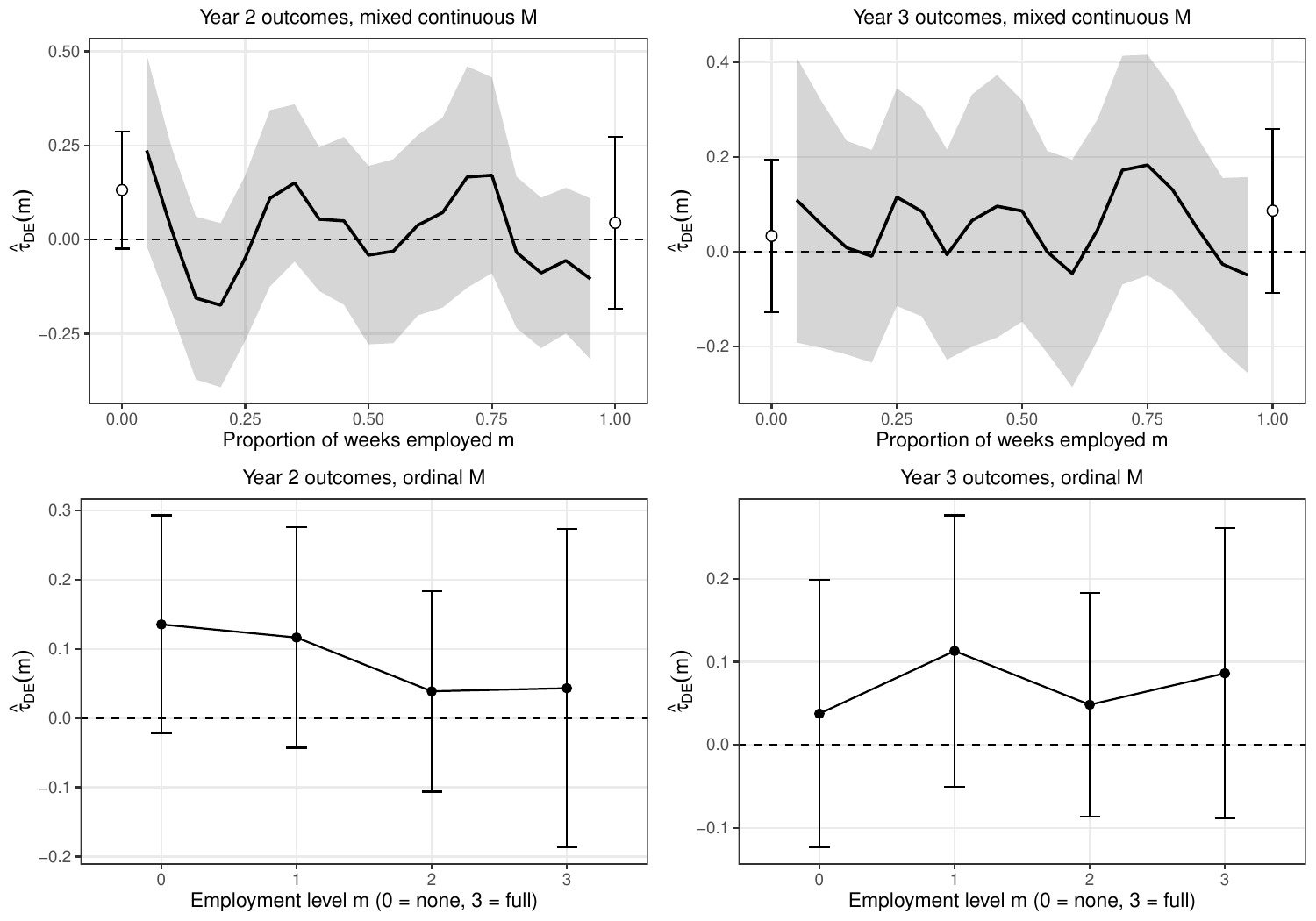}
    \caption{Controlled direct effect curves $\widehat\tau_{DE}(m)$ of year-1 training on $\log(1+\text{weekly earnings})$ in the Job Corps Study. Top: mixed continuous mediator; the curve and shaded band are the kernel-smoothed estimates and pointwise 95\% confidence intervals at interior employment levels, and the open points with error bars are the discrete estimates at the atoms $m=0$ and $m=1$. Bottom: ordinal mediator. Red segments and points indicate estimates whose 95\% confidence interval excludes zero.} \label{fig2}
\end{figure}

Figure \ref{fig2} presents the CDE curves, which are imprecise relative to the natural direct effect, as expected from the slower rate at interior points. In year 2, the CDE fluctuates around zero in the interior and is positive but insignificant at the atoms (0.13, s.e.\ 0.08 at $m=0$; 0.04, s.e.\ 0.12 at $m=1$). In year 3, the point estimates are positive over most of the range (0.08 to 0.18 for $m \in [0.70, 0.80]$; 0.09, s.e.\ 0.09 at $m=1$), but none is individually significant. The ordinal specification agrees, with the largest estimates among those employed up to half the year (0.12 and 0.11, s.e.\ 0.08). Halving or doubling the bandwidth changes the interior estimates by less than one standard error. The controlled-effect analysis is thus consistent with a modest positive direct effect roughly constant in employment intensity, which the significant natural direct effect averages over, but the data do not support strong statements about how it varies with $m$.

\section{Discussion} \label{sec:discussion}

In this paper, we studied causal mediation analysis in difference-in-differences. We showed that the total, natural indirect, and natural direct effects are identifiable in the treated group, derived their efficient influence functions, and proposed multiply robust, nonparametrically efficient estimators. The framework can be extended to staggered designs, with parallel trends conditioning on the counterfactual mediator process and sequential ignorability imposed between adjacent periods; nonparametric estimation is then demanding because each period has its own working models, and model-based methods may be a practical alternative.

The identification assumptions come at a cost, which we have tried to state plainly. The framework retains the tolerance of time-invariant treatment--outcome confounding of the standard DiD design but gives up direct verifiability of the trend assumption, retaining only the falsification test of Proposition \ref{prop:pt}. The sequential ignorability condition does not require randomization of the mediator, but it is a cross-world condition, unverifiable in principle, and it excludes treatment-induced confounders of the mediator--outcome relationship even when measured. A researcher who suspects such a confounder should instead consider randomized interventional effects, in which the mediator is drawn from its conditional distribution rather than set to its cross-world value \citep{vanderweele2017mediation}; extending them to the DiD design is a natural next step. Other approaches to relaxing sequential ignorability include structural models for the potential outcomes \citep{zheng2015causal} and instrumental variables \citep{frolich2017direct, rudolph2024using}.

On the estimation side, the pseudo-propensity score that replaces the conditional mediator density conditions on a post-treatment variable and so does not describe a data-generating mechanism; the propensity score and pseudo-propensity score models may therefore be mutually incompatible, which is why Theorem \ref{th:mr} is stated as a union over pairs of nuisance functions.

In the Job Corps Study, some participants' earnings are zero, a phenomenon known as the truncation-by-death problem \citep{zhang2009likelihood}. Zero earnings may result from self-selection in the job market, with an expected wage higher than the offered wage. Our analysis did not address this problem, and the $\log(1+Y)$ transformation does not resolve it \citep{chen2024logs}. Relatedly, when the mediator is a component of the outcome, part of the indirect effect is mechanical, and an outcome such as earnings per employed week, or hourly wages where available, is preferable.

%\section*{Acknowledgements}
%\section*{Funding information}
\section*{Conflict of interest}
The authors declare no conflict of interest.

\section*{Data availability statement}
The data that support the findings of this paper are publicly available in the R package ``causalweight''.

\section*{Supplementary material}
The supplementary material includes practical estimation procedures and proofs of theoretical results. R code to reproduce all results is available from the authors.

\bibliographystyle{apalike}
\bibliography{ref}

@article{baron1986moderator,
  title={The moderator--mediator variable distinction in social psychological research: conceptual, strategic, and statistical considerations.},
  author={Baron, Reuben M and Kenny, David A},
  journal={Journal of Personality and Social Psychology},
  volume={51},
  number={6},
  pages={1173},
  year={1986},
  publisher={American Psychological Association}
}

@article{pearl1995causal,
  title={Causal diagrams for empirical research},
  author={Pearl, Judea},
  journal={Biometrika},
  volume={82},
  number={4},
  pages={669--688},
  year={1995},
  publisher={Oxford University Press}
}

@article{heckman1997matching,
  title={Matching as an econometric evaluation estimator: Evidence from evaluating a job training programme},
  author={Heckman, James J and Ichimura, Hidehiko and Todd, Petra E},
  journal={The Review of Economic Studies},
  volume={64},
  number={4},
  pages={605--654},
  year={1997},
  publisher={Wiley-Blackwell}
}

@book{schochet2001national,
  title={National {Job Corps Study}: the impacts of {Job Corps} on participants' employment and related outcomes},
  author={Schochet, Peter Z},
  year={2001},
  publisher={US Department of Labor, Employment and Training Administration, Office of Policy and Research}
}

@article{rubin2004direct,
  title={Direct and indirect causal effects via potential outcomes},
  author={Rubin, Donald B},
  journal={Scandinavian Journal of Statistics},
  volume={31},
  number={2},
  pages={161--170},
  year={2004},
  publisher={Wiley Online Library}
}

@article{abadie2005semiparametric,
  title={Semiparametric difference-in-differences estimators},
  author={Abadie, Alberto},
  journal={The Review of Economic Studies},
  volume={72},
  number={1},
  pages={1--19},
  year={2005},
  publisher={Wiley-Blackwell}
}

@article{goetgeluk2008estimation,
  title={Estimation of controlled direct effects},
  author={Goetgeluk, Sylvie and Vansteelandt, Stijn and Goetghebeur, Els},
  journal={Journal of the Royal Statistical Society Series B: Statistical Methodology},
  volume={70},
  number={5},
  pages={1049--1066},
  year={2008},
  publisher={Oxford University Press}
}

@article{schochet2008does,
  title={Does job corps work? Impact findings from the national {Job Corps} study},
  author={Schochet, Peter Z and Burghardt, John and McConnell, Sheena},
  journal={American economic review},
  volume={98},
  number={5},
  pages={1864--1886},
  year={2008},
  publisher={American Economic Association}
}

@article{lee2009training,
  title={Training, Wages, and Sample Selection: Estimating Sharp Bounds on Treatment Effects},
  author={Lee, David S},
  journal={The Review of Economic Studies},
  volume={76},
  number={3},
  pages={1071--1102},
  year={2009},
  publisher={Review of Economic Studies Ltd}
}

@book{bickel1993efficient,
  title={Efficient and adaptive estimation for semiparametric models},
  author={Bickel, Peter J and Klaassen, Chris AJ and Bickel, Peter J and Ritov, Ya’acov and Klaassen, J and Wellner, Jon A and Ritov, YA'Acov},
  volume={4},
  year={1993},
  publisher={Springer}
}

@article{blanco2013bounds,
  title={Bounds on average and quantile treatment effects of Job Corps training on wages},
  author={Blanco, German and Flores, Carlos A and Flores-Lagunes, Alfonso},
  journal={Journal of Human Resources},
  volume={48},
  number={3},
  pages={659--701},
  year={2013},
  publisher={University of Wisconsin Press}
}

@article{flores2010learning,
  title={Learning but not earning? The impact of {Job Corps} training on Hispanic youth},
  author={Flores-Lagunes, Alfonso and Gonzalez, Arturo and Neumann, Todd},
  journal={Economic Inquiry},
  volume={48},
  number={3},
  pages={651--667},
  year={2010},
  publisher={Wiley Online Library}
}

@article{flores2012estimating,
  title={Estimating the effects of length of exposure to instruction in a training program: The case of {Job Corps}},
  author={Flores, Carlos A and Flores-Lagunes, Alfonso and Gonzalez, Arturo and Neumann, Todd C},
  journal={Review of Economics and Statistics},
  volume={94},
  number={1},
  pages={153--171},
  year={2012},
  publisher={The MIT Press}
}

@article{zhang2009likelihood,
  title={Likelihood-based analysis of causal effects of job-training programs using principal stratification},
  author={Zhang, Junni L and Rubin, Donald B and Mealli, Fabrizia},
  journal={Journal of the American Statistical Association},
  volume={104},
  number={485},
  pages={166--176},
  year={2009},
  publisher={Taylor \& Francis}
}

@article{chen2015bounds,
  title={Bounds on treatment effects in the presence of sample selection and noncompliance: the wage effects of job corps},
  author={Chen, Xuan and Flores, Carlos A},
  journal={Journal of Business \& Economic Statistics},
  volume={33},
  number={4},
  pages={523--540},
  year={2015},
  publisher={Taylor \& Francis}
}

@article{imai2010identification,
  title={Identification, Inference and Sensitivity Analysis for Causal Mediation Effects},
  author={Imai, Kosuke and Keele, Luke and Yamamoto, Teppei},
  journal={Statistical Science},
  volume={25},
  number={1},
  pages={51--71},
  year={2010}
}

@article{imai2010general,
  title={A general approach to causal mediation analysis.},
  author={Imai, Kosuke and Keele, Luke and Tingley, Dustin},
  journal={Psychological Methods},
  volume={15},
  number={4},
  pages={309},
  year={2010},
  publisher={American Psychological Association}
}

@article{zheng2015causal,
  title={Causal mediation analysis in the multilevel intervention and multicomponent mediator case},
  author={Zheng, Cheng and Zhou, Xiao-Hua},
  journal={Journal of the Royal Statistical Society Series B: Statistical Methodology},
  volume={77},
  number={3},
  pages={581--615},
  year={2015},
  publisher={Oxford University Press}
}

@article{vanderweele2017mediation,
  title={Mediation analysis with time varying exposures and mediators},
  author={VanderWeele, Tyler J and Tchetgen Tchetgen, Eric J},
  journal={Journal of the Royal Statistical Society Series B: Statistical Methodology},
  volume={79},
  number={3},
  pages={917--938},
  year={2017},
  publisher={Oxford University Press}
}

@article{frolich2017direct,
  title={Direct and indirect treatment effects--causal chains and mediation analysis with instrumental variables},
  author={Fr{\"o}lich, Markus and Huber, Martin},
  journal={Journal of the Royal Statistical Society Series B: Statistical Methodology},
  volume={79},
  number={5},
  pages={1645--1666},
  year={2017},
  publisher={Oxford University Press}
}

@article{chernozhukov2018double,
  title={Double/debiased machine learning for treatment and structural parameters},
  author={Chernozhukov, Victor and Chetverikov, Denis and Demirer, Mert and Duflo, Esther and Hansen, Christian and Newey, Whitney and Robins, James},
  journal={The Econometrics Journal},
  volume={21},
  number={1},
  pages={C1--C68},
  year={2018},
  publisher={JSTOR}
}

@article{deuchert2019direct,
  title={Direct and indirect effects based on difference-in-differences with an application to political preferences following the Vietnam draft lottery},
  author={Deuchert, Eva and Huber, Martin and Schelker, Mark},
  journal={Journal of Business \& Economic Statistics},
  volume={37},
  number={4},
  pages={710--720},
  year={2019},
  publisher={Taylor \& Francis}
}

@article{sant2020doubly,
  title={Doubly robust difference-in-differences estimators},
  author={Sant’Anna, Pedro HC and Zhao, Jun},
  journal={Journal of Econometrics},
  volume={219},
  number={1},
  pages={101--122},
  year={2020},
  publisher={Elsevier}
}

@article{callaway2021difference,
  title={Difference-in-differences with multiple time periods},
  author={Callaway, Brantly and Sant’Anna, Pedro HC},
  journal={Journal of Econometrics},
  volume={225},
  number={2},
  pages={200--230},
  year={2021},
  publisher={Elsevier}
}

@article{huber2022direct,
  title={Direct and indirect effects based on changes-in-changes},
  author={Huber, Martin and Schelker, Mark and Strittmatter, Anthony},
  journal={Journal of Business \& Economic Statistics},
  volume={40},
  number={1},
  pages={432--443},
  year={2022},
  publisher={Taylor \& Francis}
}

@incollection{vaart2023empirical,
  title={Empirical processes},
  author={{van der Vaart}, AW and Wellner, Jon A},
  booktitle={Weak Convergence and Empirical Processes: With Applications to Statistics},
  pages={127--384},
  year={2023},
  publisher={Springer}
}

@article{zivich2023introducing,
  title={Introducing proximal causal inference for epidemiologists},
  author={Zivich, Paul N and Cole, Stephen R and Edwards, Jessie K and Mulholland, Grace E and Shook-Sa, Bonnie E and Tchetgen Tchetgen, Eric J},
  journal={American Journal of Epidemiology},
  volume={192},
  number={7},
  pages={1224--1227},
  year={2023},
  publisher={Oxford University Press}
}

@article{rudolph2024using,
  title={Using instrumental variables to address unmeasured confounding in causal mediation analysis},
  author={Rudolph, Kara E and Williams, Nicholas and D{\'\i}az, Iv{\'a}n},
  journal={Biometrics},
  volume={80},
  number={1},
  pages={ujad037},
  year={2024},
  publisher={Oxford University Press}
}

@article{deng2025improved,
  title={Improved two-period difference-in-differences by targeted estimation},
  author={Deng, Yuhao and Zhang, Tao and Peng, Xiang and Liu, Qinqing},
  journal={Economics Letters},
  pages={112600},
  year={2025},
  publisher={Elsevier}
}

@article{hsia2025causal,
  title={Causal mediation analysis for difference-in-difference design and panel data},
  author={Hsia, Pei-Hsuan and Tai, An-Shun and Kao, Chu-Lan Michael and Lin, Yu-Hsuan and Lin, Sheng-Hsuan},
  journal={Epidemiologic Methods},
  volume={14},
  number={1},
  pages={20240025},
  year={2025},
  publisher={De Gruyter}
}

@article{blackwell2025estimating,
  title={Estimating controlled direct effects with panel data: an application to reducing support for discriminatory policies},
  author={Blackwell, Matthew and Glynn, Adam N and Hilbig, Hanno and Phillips, Connor Halloran},
  journal={Journal of the Royal Statistical Society Series A: Statistics in Society},
  pages={qnaf042},
  year={2025},
  publisher={Oxford University Press UK}
}

@article{chang2020double,
  title={Double/debiased machine learning for difference-in-differences models},
  author={Chang, Neng-Chieh},
  journal={The Econometrics Journal},
  volume={23},
  number={2},
  pages={177--191},
  year={2020},
  publisher={Oxford University Press}
}

@book{silverman2018density,
  title={Density estimation for statistics and data analysis},
  author={Silverman, Bernard W},
  year={2018},
  publisher={Routledge}
}

@article{huber2026difference,
  title={Difference-in-differences for mediation analysis using double machine learning},
  author={Huber, Martin and Oberh{\"a}nsli, Sarina Joy},
  journal={arXiv preprint arXiv:2602.23877},
  year={2026}
}

@article{chen2024logs,
  title={Logs with zeros? Some problems and solutions},
  author={Chen, Jiafeng and Roth, Jonathan},
  journal={The Quarterly Journal of Economics},
  volume={139},
  number={2},
  pages={891--936},
  year={2024},
  publisher={Oxford University Press}
}

@article{joffe2009related,
  title={Related causal frameworks for surrogate outcomes},
  author={Joffe, Marshall M and Greene, Tom},
  journal={Biometrics},
  volume={65},
  number={2},
  pages={530--538},
  year={2009},
  publisher={Wiley}
}

@article{kennedy2017non,
  title={Non-parametric methods for doubly robust estimation of continuous treatment effects},
  author={Kennedy, Edward H and Ma, Zongming and McHugh, Matthew D and Small, Dylan S},
  journal={Journal of the Royal Statistical Society Series B: Statistical Methodology},
  volume={79},
  number={4},
  pages={1229--1245},
  year={2017},
  publisher={Oxford University Press}
}

@article{chernozhukov2022automatic,
  title={Automatic debiased machine learning of causal and structural effects},
  author={Chernozhukov, Victor and Newey, Whitney K and Singh, Rahul},
  journal={Econometrica},
  volume={90},
  number={3},
  pages={967--1027},
  year={2022},
  publisher={Wiley Online Library}
}

\includepdf[pages=1-31]{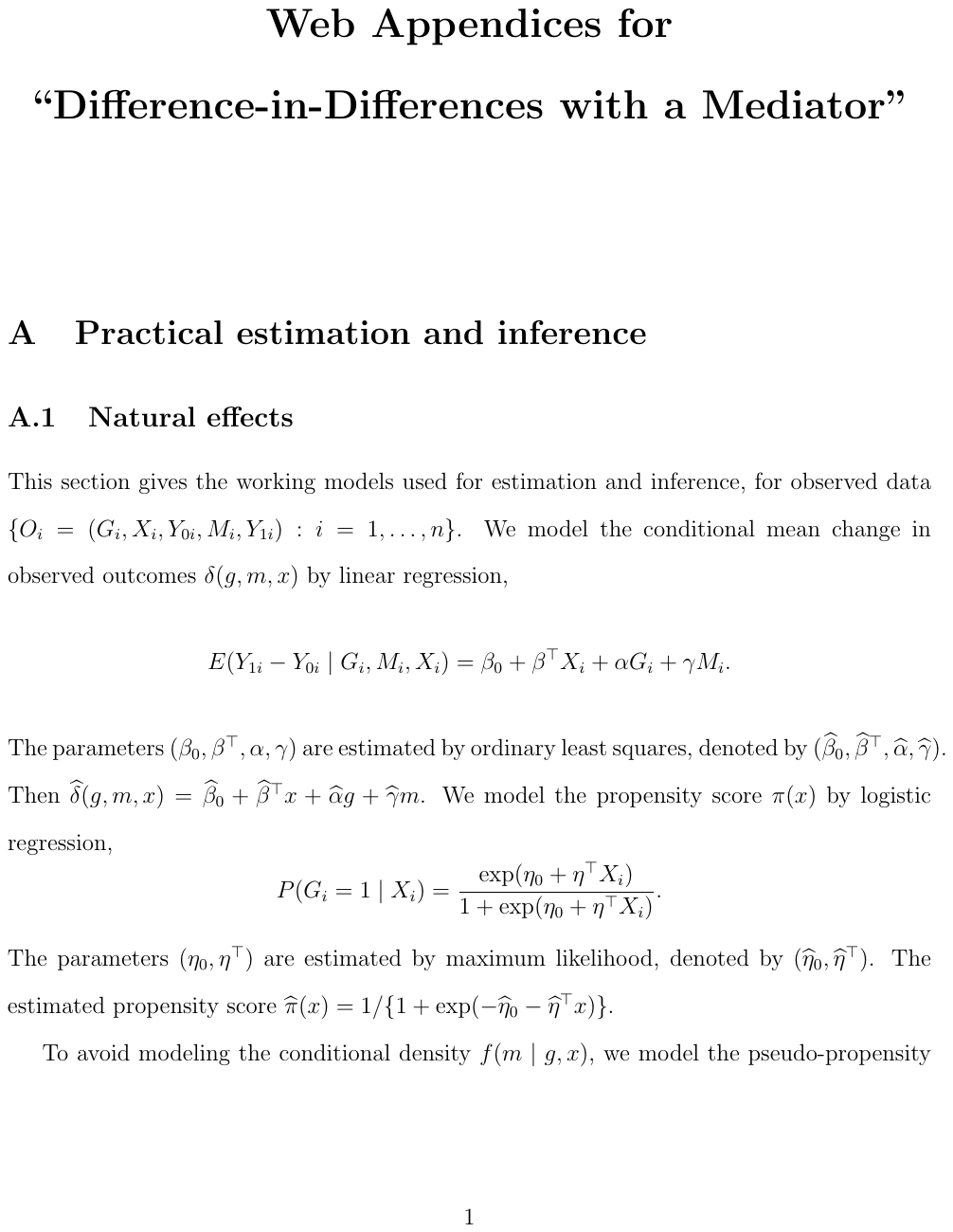}

\end{document}